\input harvmac
\def\IB{\relax\hbox{$\inbar\kern-.3em{\rm B}$}}
\def\IC{\relax\hbox{$\inbar\kern-.3em{\rm C}$}}
\def\ID{\relax\hbox{$\inbar\kern-.3em{\rm D}$}}
\def\IE{\relax\hbox{$\inbar\kern-.3em{\rm E}$}}
\def\IF{\relax\hbox{$\inbar\kern-.3em{\rm F}$}}
\def\IG{\relax\hbox{$\inbar\kern-.3em{\rm G}$}}
\def\IGa{\relax\hbox{${\rm I}\kern-.18em\Gamma$}}
\def\IH{\relax{\rm I\kern-.18em H}}
\def\IK{\relax{\rm I\kern-.18em K}}
\def\IL{\relax{\rm I\kern-.18em L}}
\def\IP{\relax{\rm I\kern-.18em P}}
\def\IR{\relax{\rm I\kern-.18em R}}
\def\IZ{\relax\ifmmode\mathchoice
{\hbox{\cmss Z\kern-.4em Z}}{\hbox{\cmss Z\kern-.4em Z}}
{\lower.9pt\hbox{\cmsss Z\kern-.4em Z}}
{\lower1.2pt\hbox{\cmsss Z\kern-.4em Z}}\else{\cmss
Z\kern-.4em
Z}\fi}

\def\II{\relax{\rm I\kern-.18em I}}


\def\CD {{\cal D}}

\def\CF {{\cal F}}

\def\CH {{\cal H}}

\def\CK {{\cal K}}

\def\CN {{\cal N}}
\def\CO {{\cal O}}

\def\CS {{\cal S}}

\def\CU {{\cal U}}
\def\CV {{\cal V}}

\def\CZ {{\cal Z}}

\def\p{\partial}


\def\zb {\bar{z}}


\def\Tr{{\rm Tr}}

\def\p{\partial}

\def\np{\nabla_{\partial}}

\def\Det{{\rm Det}}

\def\inbar{\,\vrule height1.5ex width.4pt depth0pt}
\font\cmss=cmss10 \font\cmsss=cmss10 at 7pt

\def\a{\alpha}

\def\b{\beta}
\def\g{\gamma}

\def\e{\epsilon}
\def\m{\mu}

\def\l{\lambda}
\def\ve{\varepsilon}
\def\z{\zeta}

\def\p{\partial}

\def\up{\upsilon}

\def\R{\relax{\rm I\kern-.18em R}}
\font\cmss=cmss10 \font\cmsss=cmss10 at 7pt
\def\Z{\relax\ifmmode\mathchoice
{\hbox{\cmss Z\kern-.4em Z}}{\hbox{\cmss Z\kern-.4em Z}}
{\lower.9pt\hbox{\cmsss Z\kern-.4em Z}}
{\lower1.2pt\hbox{\cmsss Z\kern-.4em Z}}\else{\cmss
Z\kern-.4em
Z}\fi}

\def\np{{\it Nucl. Phys. }}

\def\hf{{1\over 2}}

\lref\hoppe{J.~Hoppe,  "Quantum theory of a massless
relativistic surface ..."
Elementary Particle Research Journal (Kyoto) 80 (1989)
}
\lref\gerasimov{A. Gerasimov, ``Localization in
GWZW and Verlinde formula,'' hepth/9305090}
\lref\kirwan{F.~Kirwan, ``Cohomology of quotients in
symplectic
and algebraic geometry'', Math. Notes, Princeton
University Press,
1985}
\lref\krwjffr{L.C.~Jeffrey, F.C.~Kirwan
``Localization for nonabelian group actions'',
alg-geom/9307001}
\lref\givental{A.B.~Givental,
``Equivariant Gromov - Witten Invariants'',
alg-geom/9603021.}
\lref\Witdgt{ E.~ Witten,  Commun. Math. Phys.  141  (1991) 153.}

\lref\morozov{A. Marshakov at al., Phys. Lett.
265B (1991) 99.}
\lref\SeWi{N. Seiberg and E. Witten,  
Nucl. Phys. B426 (1994) 19-52 (and erratum - ibid. B430 (1994)
485-486 ), 
Nucl. Phys.  B 431 (1994) 484-550.}

\lref\bfss{T. Banks, W. Fischler, S. Shenker, and L. Susskind, Phys. Rev. 55 (1997) 112.}%
 \lref\polchin{S. Chaudhuri, C. Johnson, and J. Polchinski,
``Notes on D-branes,'' hep-th/9602052; J. Polchinski,
``TASI Lectures on D-branes,'' hep-th/9611050.}
\lref\ginz{V.~Ginzburg, R.~Besrukavnikov, to appear}
   \lref\horostrom{G.T. Horowitz and A. Strominger,
``Black strings and $p$-branes,''
Nucl. Phys. {\bf B} 360(1991) 197.}
\lref\nicolai{W.~Krauth, H.~Nicolai, M.~Staudacher,
``Monte Carlo Approach to M-theory'', hep-th/9803117.}
\lref\townsend{P. Townsend, ``The eleven dimensional
supermembrane
revisited,'' hep-th/9501068}
\lref\porrati{M.~Porrati, A.~Rozenberg,
``Bound States at Threshold in Supersymmetric Quantum
Mechanics'',
hep-th/9708119}
\lref\ikkt{N. Ishibashi, H. Kawai, Y. Kitazawa, and A.
Tsuchiya,
``A large $N$ reduced model as superstring,'' hep-th/9612115;
Nucl. Phys. {\bf B} 498 (1997)467.}

\lref\dsl{E. Brezin and V. Kazakov, Phys. Lett. 236B (1990) 144\semi
 M. Douglas and
S. Shenker, Nucl.Phys. B 335 (1990) \semi
D. Gross and A.A. Migdal, Phys. Rev. Lett. 64 (1990) 127.}

\lref\witdyn{E. Witten, ``String theory dynamics
in various dimensions,''
hep-th/9503124, Nucl. Phys. {\bf B} 443 (1995) 85-126.}
\lref\WitDonagi{R.~ Donagi, E.~ Witten,
``Supersymmetric Yang-Mills Theory and
Integrable Systems'', hep-th/9510101, Nucl. Phys.{\bf B}
460 (1996)
299-334.}
\lref\Witfeb{E.~ Witten, ``Supersymmetric Yang-Mills
Theory On A
Four-Manifold,''  hep-th/9403195; J. Math. Phys. {\bf 35}
(1994)
5101.}
\lref\Witr{E.~ Witten, ``Introduction to Cohomological Field
Theories",
Lectures at Workshop on Topological Methods in Physics,
Trieste,
Italy,
Jun 11-25, 1990, Int. J. Mod. Phys. {\bf A6} (1991) 2775.}
\lref\Witgrav{E.~ Witten, ``Topological Gravity'',
Phys.Lett.206B:601, 1988.}
\lref\witaffl{I. ~ Affleck, J.A.~ Harvey and E.~ Witten,
        ``Instantons and (Super)Symmetry Breaking
        in $2+1$ Dimensions'', Nucl. Phys. {\bf B}206
(1982) 413.}
\lref\wittabl{E.~ Witten,  ``On $S$-Duality in Abelian Gauge
Theory,''
hep-th/9505186; Selecta Mathematica {\bf 1} (1995) 383.}
\lref\wittgr{E.~ Witten, ``The Verlinde Algebra And The
Cohomology Of
The Grassmannian'',  hep-th/9312104.}
\lref\wittenwzw{E. Witten, ``Nonabelian bosonization in
two dimensions,'' Commun. Math. Phys. {\bf 92} (1984)455 }
\lref\witgrsm{E. Witten, ``Quantum field theory,
grassmannians and algebraic curves,''
Commun.Math.Phys.113:529,1988.}
\lref\wittjones{E. Witten, ``Quantum field theory and the
Jones
polynomial,'' Commun.  Math. Phys., 121 (1989) 351. }
\lref\witttft{E.~ Witten, ``Topological Quantum Field Theory",
Commun. Math. Phys. {\bf 117} (1988) 353.}
\lref\wittmon{E.~ Witten, ``Monopoles and Four-Manifolds'',
hep-th/9411102}
 
\lref\CHSW{P.~Candelas, G.~Horowitz, A.~Strominger and
E.~Witten,
``Vacuum Configurations for Superstrings,'' {\it Nucl.
Phys.}  
B258 (1985) 46.}

\lref\AandB{E.~Witten, in ``Proceedings of the Conference
on Mirror
Symmetry",
MSRI (1991).}

\lref\phases{E.~Witten, ``Phases of N=2 Theories in Two
Dimensions",
Nucl. Phys. {\bf B403} (1993) 159, hep-th/9301042.}
\lref\WitMin{E.~Witten,
``On the Landau-Ginzburg Description of N=2 Minimal Models,''
IASSNS-HEP-93/10, hep-th/9304026.}

\lref\wittenwzw{E. Witten, ``Nonabelian bosonization in
two dimensions,'' Commun. Math. Phys. {\bf 92} (1984) 455 .}
\lref\grssmm{E. Witten, ``Quantum field theory,
grassmannians and algebraic curves,''
Commun.Math.Phys.113 (1988) 529.}
\lref\wittjones{E. Witten, ``Quantum field theory and the
Jones
polynomial,'' Commun.  Math. Phys., 121 (1989) 351. }
\lref\wittentft{E.~ Witten, ``Topological Quantum Field
Theory",
Commun. Math. Phys. {\bf 117} (1988) 353.}
\lref\Witr{E.~ Witten, ``Introduction to Cohomological Field
Theories",
Lectures at Workshop on Topological Methods in Physics,
Trieste,
Italy,
Jun 11-25, 1990, Int. J. Mod. Phys. {\bf A6} (1991) 2775.}
\lref\wittabl{E. Witten,  ``On S-Duality in Abelian Gauge
Theory,''
hep-th/9505186}
\lref\witbound{E.~Witten,  Nucl. Phys. B460 (1996) 335.}
\lref\witconst{E.~Witten, ``Constraints on supersymmetry
breaking'',
Nucl. Phys. {\bf B}202 (1982) 253.}
 
\lref\kristjansen{ B. Eynard and  C. Kristjansen, 
Nucl. Phys. B516 (1998) 529}
\lref\gaudin{M. Gaudin and I. Kostov, Phys. Lett. B220 (1989) 200.}
\lref\kostovOn{I. Kostov, Mod. Phys. Lett.  A4 (1989)
217.}
\lref\kostovOnKdV{I. Kostov, Nucl. Phys. B (Proc. Suppl.) 45A (1996) 13, 
hep-th/9509124.}
 \lref\kostov{I. K. Kostov, Nucl.Phys. {\bf B}376 (1992)  
539\semi
I. K. Kostov and  M. Staudacher,
Nucl.Phys. {\bf B}384 (1992) 459.}
\lref\slavnov{See, e.g. V. E.~Korepin and  N. A.~Slavnov, ``The
determinant representation for quantum correlation  
functions of the
sinh-Gordon model'', hep-th/9801046, and
references therein.}
\lref\vw{C.~Vafa, E.~Witten, ``A strong coupling test of
$S$-duality'', hep-th/9408074;
Nucl. Phys. B 431 (1994) 3-77}
\lref\grojn{I. Grojnowski, ``Instantons and
affine algebras I: the Hilbert scheme and
vertex operators,'' alg-geom/9506020}
\lref\gr{G.~Gibbons, P.~Rychenkova ``hyperkahler quotient
construction
of BPS Monopole moduli space'', hep-th/9608085}
\lref\dvafa{C.~Vafa, ``Instantons on D-branes'',
hep-th/9512078,
Nucl. Phys. B463 (1996) 435-442}
\lref\atbott{M.~Atiyah, R.~Bott, ``The Moment Map And
Equivariant Cohomology'', Topology {\bf 23} (1984) 1-28}
\lref\atbotti{M.~Atiyah, R.~Bott, ``The Yang-Mills
Equations Over
Riemann Surfaces'', Phil. Trans. R.Soc. London A {\bf
308}, 523-615
(1982)}
\lref\marty{M.~Claudson and  M.B.~Halpern,   
Nucl. Phys. B 250(1985)  689.}
\lref\polchin{S. Chaudhuri, C. Johnson, and J. Polchinski,
``Notes on D-branes,'' hep-th/9602052; J. Polchinski,
``TASI Lectures on D-branes,'' hep-th/9611050}
\lref\ginz{V.~Ginzburg, R.~Besrukavnikov, to appear}
\lref\grn{M.~Green and  M.~Gutperle, ``$D$-particle bound
states and the
$D$-instanton measure'',
hep-th/9711107.}
\lref\sav{S.~Sethi and  M.~Stern, ``$D$-brane bound states
redux,''
hep-th/9705046.}
\lref\pyi{P.~Yi,
  Nucl. Phys. B 505 (1997) 307.}
\lref\higgs{G.~Moore, N.~Nekrasov and  S.~Shatashvili,
``Integrating over
Higgs branches'',
hep-th/9712241.}
\lref\kato{S. Hirano and  M. Kato, Prog.~Theor.~Phys. 98 (1997) 1371.}
\lref\index{G.~Moore, N.~Nekrasov, S.~Shatashvili,
``D-particle bound
states and generalized instantons'',
HUTP-98/A008, ITEP-TH-8/98,
hep-th/9803265.}
\lref\horostrom{G.T. Horowitz and A. Strominger,
Nucl. Phys. B360 (1991) 197.}
\lref\nicolai{W.~Krauth, H.~Nicolai, M.~Staudacher,
``Monte Carlo Approach to M-theory'', hep-th/9803117}
\lref\manin{Yu.~Manin, ``Generating functions in algebraic
geometry and sums over trees'', alg-geom/9407005}
\lref\estring{J.A.~Minahan, D.~Nemeschansky, C.~Vafa, N.
P.~Warner,
``E-Strings and $\CN=4$ Topological Yang-Mills Theories'',
hep-th/9802168}
\lref\townsend{P. Townsend, ``The eleven dimensional
supermembrane
revisited,'' hep-th/9501068.}
\lref\porrati{M.~Porrati, A.~Rozenberg,
``Bound States at Threshold in Supersymmetric Quantum
Mechanics'',
hep-th/9708119}
\lref\ikkt{N. Ishibashi, H. Kawai, Y. Kitazawa and A.
Tsuchiya, Nucl. Phys.  B498 (1997) 467.}
\lref\nikitathesis{N.~Nekrasov, ``Four dimensional holomorphic
theories'', PhD Thesis,
Princeton, 1995.}
 \lref\jimiwa{
 M. Jimbo and T. Miwa, "Solitons and infinite dimensional Lie
algebras",
{\it RIMS } Vol. 19 (1983) 943-1001, and references therein.}
 \lref\hirota{R. Hirota, Phys. Rev. Lett. 27 (1971) 1192.}
\lref\mehta{M. L. Mehta,  {\it Random Matrices}, second 
edition
(Academic Press, New York, 1990).}
 \lref\izii{C. Itzykson and J.-B. Zuber, J. Math. Phys.  
21 (1980)
411.}
 \lref\zamsal{P. Fendley and H. Saleur, \np B 388 (1992) 609,
 Al. B. Zamolodchikov, \np B 432 [FS] (1994) 427.}
\lref\trw{ D. Bernard and A. LeClair, \np {\bf B} 426  
(1994) 534;
 C. Tracy and H. Widom, ``Fredholm determinants and
the mKdV/sinh-Gordon hierarchies'', solv-int/9506006, \semi
S.~Kakei, ``Toda
Lattice
Hierarchy and Zamlodchikov conjecture'', solv-int/9510006.}
\lref\DS{V.~Drinfeld and V.~Sokolov, {\it. Itogi Nauki i
Techniki} 24
(1984)
81,
{\it Docl. Akad. Nauk. SSSR} 258 (1981) 1.}
\lref\DFK{P. Di Francesco and D. Kutasov, \np B 342
(1990) 589.}
\lref\KW{V. Kac and M. Wakimoto, ``Exceptional hierarchies
of soliton
equations'', {\it Proceedings of Symposia in Pure  
Mathematics}, 49
(1989) 191}
\lref\BIPZ{E. Brezin, C. Itzykson, G. Parisi  and  J.B. Zuber, Commun.
Math. Phys. 59 (1978) 35.}
\lref\IvanRazl{I. Kostov, Proceedings of the $2^{\rm nd}$
Bulgarian Workshop  "New trends in quantum field theory", Razlog, 
Bulgaria, 28 August - 1 September 1995, hep-th/9602117.}

\Title{
\vbox{\baselineskip12pt
\hbox{HUTP-98/A051}
\hbox{ITEP-TH-35/98}
\hbox{CERN-TH/98-302}
\hbox{SPHT-t98/102}
\hbox{LPTENS-98/40}
}}
{\vbox{ \centerline
{D-particles, matrix integrals  and KP hierarchy  }}}
\footnote{}{kazakov@peterpan.ens.fr, ivan.kostov@cern.ch,
nikita@string.harvard.edu }

\centerline{Vladimir A.  Kazakov$^1$\footnote{$^\diamond$}{Research
supported in part by European contract TMR ERBFMRXCT960012},
Ivan K. Kostov $^{2 \diamond}$\footnote{$ ^\dagger$}{Member of CNRS}
 and Nikita Nekrasov$^3$}

\vskip .3in

{\it $^1$  Laboratoire de Physique Th\'eorique de l'Ecole
Normale Sup\'erieure\footnote{$ ^\ast $}{
Unit\'e Propre du
Centre National de la Recherche Scientifique,
associ\'ee \`a l'Ecole Normale Sup\'erieure et \`a
l'Universit\'e de Paris-Sud.},}

{\ \ \ \it  75231 Paris, France}

{\it $^2$ CERN, Theory Division, 1211 Geneva 23, Switzerland , and}

{\ \ \ \it C.E.A. - Saclay, Service de
Physique Th\'eorique,  F-91191 Gif-Sur-Yvette, France}

{\it $^3$ Institute
 of Theoretical and Experimental
Physics, 117259, Moscow, Russia, and}

{\ \ \ \it  Lyman Laboratory of Physics,
Harvard University, Cambridge, MA 02138, USA}

\vskip .2in

We study the regularized correlation functions of the light-like
 coordinate operators in the reduction to zero dimensions of the
 matrix model describing $D$-particles in four dimensions. We
 investigate in great detail the related matrix model originally
 proposed and solved in the planar limit by J.~Hoppe. It also gives
 the solution of the problem of 3-coloring of planar graphs. We find
 interesting strong/weak 't Hooft coupling dependence.  The partition
 function of the grand canonical ensemble turns out to be a
 tau-function of KP hierarchy.  As an illustration of the method we
 present a new derivation of the large-$N$ and double-scaling limits
 of the one-matrix model with cubic potential.  \bigskip
\Date{10/98}

\newsec{Introduction }

Matrix models describing the behaviour of $Dp$-branes originate in the
observation of Witten \witbound\ that the massless modes propagating
along the world volume of $N$ coincident $D$-branes are those of the
supersymmetric Yang-Mills theory, obtained by the dimensional
reductions of the $d=10$ $\CN=1$ theory down to $p+1$ space-time
dimensions.

In various compactifications of string theory one
encounters the
nearly massless non-perturbative particles, obtained by
wrapping
the $Dp$-branes around vanishing $p$-cycles inside the
internal
Calabi-Yau
manifold.
Even in ten-dimensional Type IIA string theory there are
solitonic
particles \horostrom, which are represented by certain
black holes
in the effective supergravity and are interpreted as
Kaluza-Klein
modes of the graviton multiplet in the compactification of
$M$-theory on a circle \refs{\townsend , \bfss}. Of course, these
particles are no longer massless.

Despite the variety of mechanisms by which such objects
appear, their
internal
description at low energies proves to be rather simple.
In fact, if
$N$ such particles in $d$ space-time dimensions
are close to one another then  their dynamics
is described by the dimensional reduction of $\CN=1$
super-Yang-Mills
theory from $d$ down to $0+1$ dimensions
(first studied a
long time ago for
different reasons in \marty). The
degrees of freedom in such quantum mechanics are
represented by
$U(N)$
matrices $X^{i}$, $i = 1, \ldots, d$,
together with the gauge field $A_{t}$ and their fermionic
partners.

Although the exact computations in quantum mechanics of
interacting
particles
are rarely possible, the supersymmetry allows one to get
some exact answers. In this paper we shall concentrate
on the correlation functions of the light-like coordinate
operator.
To state  more precisely what  we mean by that, let us
consider the
quantum mechanics with periodic time $t \sim t
+2\pi \beta$
and with periodic boundary
conditions on fermions. In this case one
can show that the observable
$$
\CO_{R} = {\Tr}_{R} P\exp\oint dt \left( A_{t} +  X^{3}
\right)
$$
commutes with some of the supercharges (of course the
choice of
$X^{3}$
is arbitrary).
In the  limit $\beta \to 0$ (and after Wick rotation)
the computations in the quantum mechanics
reduce to the finite-dimensional integrals, where $A_{t}$
becomes the
$0$-th
matrix $X_0=-iX_4$.
Then the observables $\CO_{R}$ can be expanded in
$$
{\Tr} \left( X^{+} \right)^{l}, \quad X^{+} = X^{3} + i X^{4}.
$$

The paper is organized as follows. We are going to study the case
$d=4$ in great detail.  We derive the determinant representation for
the regularized generating function of the correlators of ${\Tr}
\left( X^{+}
\right)^{l}$
 and show that it obeys Hirota bilinear identities (when working with
fixed chemical potential, e.g. in the grand canonical ensemble). Then
we concentrate on the operators $\left( {\Tr} \left( X^{+} \right)^{2}
\right)^{l}$
and derive the asymptotics for the generating function in certain
limits.  We then briefly discuss $d=6,10$ cases. Then we proceed with
the direct attack on the $d=4$ integral for fixed but large $N$, using
the saddle-point techniques, and derive interesting asymptotics both
in the strong and in the weak coupling limit. In the weak coupling
limit, we obtain agreement with the planar graph expansion. In the
strong 't Hooft coupling limit we get the agreement with the
predictions from KP hierarchy.  In the first appendix we check the
strong coupling asymptotics by the direct quasiclassical calculation
of the matrix integral.  In the second appendix we demonstrate our
method based on the KP differential equation on the example of the
usual one matrix model.

In the bulk of the paper we use the notation $\phi \equiv
X^{+},
\bar\phi =
X^{-}$. We also denote by $Z, \, F = {\rm log} Z$ the partition
function and the free energy at fixed particle number $N$ and by $\CZ,
\, \CF = {\rm log}\CZ$ the corresponding quantities at the fixed
chemical potential $\mu$.

\newsec{Supersymmetric matrix integrals}

\subsec{Theory with four supercharges}

The dimensional reduction of  the
$\CN=1$ SYM from $d=4$ dimensions down
to zero dimensions would produce a matrix model with $4$
bosonic
matrices $X_{\mu}$,
$\mu = 1,2,3,4$ and $2$ complex fermionic matrices
$\lambda_{a}$,
$a = 1,2$. All matrices are in the adjoint representation
of the
gauge group $G$, which we will take to be either $U(N)$ or
$SU(N)/{\IZ}_{N}$.
The matrix integral has the form:
\eqn\mit{{1\over{{\rm Vol}(G)}} \int d X d \lambda
\exp \left(
{\half}
\sum_{\mu < \nu} {\Tr} [ X_{\mu}, X_{\nu} ]^{2}  + {\Tr}
\bar\lambda_{\dot a}
\sigma_{\mu}^{a \dot a} [ X_{\mu}, \lambda_{a}] \right), }
where $\sigma_{\mu} = ( 1, \sigma_{i}), i=1,2,3$, 
 are the Pauli
matrices.
The two complex fermions $\lambda$ can be viewed as four real
fermions, 
 which we denote by $\chi, \eta, \psi_{\alpha}$, $\alpha = 1,2$:
$$
\lambda_{1} = \half \left( \eta - i \chi \right), \quad
\lambda_{2} =
\half
 \left( \psi_{1} + i \psi_{2} \right)
$$
and
$$
\bar \lambda_{\dot a} = \sigma_{2}^{a\dot a} \lambda_{a}^{*}.
$$
We also redefine the bosonic
matrices as:
\eqn\rdfn{\phi = {1\over{\sqrt{2}}} \left( X_{3} + i
X_{4} \right),\ \ \ \
\bar \phi = {1\over{\sqrt{2}}} \left( X_{3} - i X_{4}
\right) }
and introduce an auxiliary  bosonic field $H$ (also in
the adjoint).
Then the integral in \mit\ becomes:
\eqn\nrprs{\eqalign{& \int {{dX_{\alpha}d\psi_{\alpha}d\chi dH
d\bar\phi d\eta d\phi}\over{{\rm Vol}(G)}}
\exp (- S) \cr
S = & \left(  i {\Tr} H s + {\half} {\Tr} H^{2} + {\Tr}
[X_{\alpha}, \phi] [X_{\alpha} , \bar\phi]  +
{\half}{\Tr} [\phi, \bar\phi]^{2} + \ldots \right)\cr}}
where $s = [ X_{1}, X_{2}]$ and
 $\ldots$ represent the fermionic terms that are
reconstructed using
the following nilpotent symmetry of \nrprs:
\eqn\nilp{\eqalign{\delta X_{\alpha} = \psi_{\alpha},
\quad & \quad
\delta \psi_{\alpha} = [ \phi, X_{\alpha}] \cr
\delta \bar\phi = \eta, \quad & \quad \delta \eta = [
\phi, \bar\phi]
\cr
\delta \chi = H, \quad & \quad \delta H = [ \phi, \chi] \cr
\qquad \qquad \delta \phi & = 0. \qquad \qquad \cr}}
The symmetry $\delta$ squares to the gauge transformation
generated
by $\phi$, hence it is nilpotent on the gauge-invariant
quantities.
This symmetry was formally studied in \kato\ in order to
apply it to
the model of
\ikkt ; it was powerfully exploited in \index\ in the
problem of
computing the 
Witten index in certain quantum mechanical systems (first
studied in the 
two-particle
case in \refs{ \pyi - \sav}, see also \grn).
The action of the matrix integral \nrprs\
 is
$\delta$-exact; in fact it may be written as
\eqn\actn{S = \delta \left( - i{\Tr}  \chi s - {\half}
{\Tr} \chi H -
\sum_{\alpha} {\Tr} \psi_{\alpha} [X_{\alpha}, \bar\phi] -
{\half} {\Tr} \eta [\phi, \bar\phi] \right).}

Now we proceed to reducing the integral \nrprs\ to
an integral with respect to the single matrix variable $\phi$.
 The strategy is known for some time
\Witdgt ,  and it consists of two steps.

If the action is perturbed by the expression $\delta (
\ldots)$, which has a
  nice behaviour at infinity, then the integral
should not change, 
which can be shown by doing an integration by parts.
Consider the modification of the action $S$ by the term
$$
S \to S + i  \delta R, 
$$
with
\eqn\pertu{R = {{\kappa_1}\over{2}}
\varepsilon^{\alpha\beta} {\Tr}
\psi_{\alpha} X_{\beta} +
\kappa_2 {\Tr} \chi \bar\phi . }
This perturbation makes the integral \mit\ localized near
the zeros
of $H, \bar\phi, \chi, \eta$ in  the limit of large
$\kappa_2$, which
can be shown by the saddle-point approximation.
It reduces the  integral \mit\ to a simpler one
\eqn\rdin{\int {{d\phi d X_{\alpha} d \psi_{\alpha}}\over{{\rm
Vol}(G)}}
\exp( i\kappa_1 {\Tr} \phi [ X_{1}, X_{2} ] + \kappa_1
\psi_{1}
\psi_{2}).}

The behaviour of the integrand at large values of
$\phi$ is still not good enough.
To make it better behaved we modify the transformation
$\delta$.
The current $\delta$ is designed to respect the ordinary gauge
invariance.
In particular $\delta^{2}$ is a gauge transformation
generated by
$\phi$.
We wish to invoke yet another symmetry of the integral
in \mit,  which
is the global group $U(1)_{\epsilon}$
acting on the matrices $X_{\alpha}, \psi_{\alpha}$
via the rotations:
\eqn\glob{e^{i\theta} : X_{1} + i X_{2} \mapsto
e^{i\theta} \left(
X_{1} +
i X_{2} \right). }
The rest of the fields are invariant under this
$U(1)_{\epsilon}$
group action.
Let us denote the generator of this group by $\epsilon$.
Then the new
supercharge
$\delta$ acts as follows:
\eqn\mnilp{\eqalign{\delta X_{\alpha} = \psi_{\alpha},
\quad & \quad
\delta \psi_{\alpha} = [ \phi, X_{\alpha}]  + i \epsilon \cdot
\varepsilon^{\alpha\beta} X_{\beta} \cr
\delta \bar\phi = \eta, \quad & \quad \delta \eta = [
\phi, \bar\phi]
\cr
\delta \chi = H, \quad & \quad \delta H = [ \phi, \chi] \cr
\qquad \qquad \delta \phi & = 0. \qquad \qquad \cr}}

The integral \mit\
has another  $U(1)$ symmetry (called the ghost number
$U(1)_{\rm gh}$)
under which
$\delta$ has charge $+1$, the bosons $X_{\alpha}, H$ have
charge $0$,
the fermions $\psi_{\alpha}$ have charge $+1$, the
fermions $\chi,
\eta$
have charge $-1$, and the bosons $\phi$ and $\bar\phi$ have
charges $+2$ and $-2$, respectively. The measure and the action
have the  over-all charge $0$. The modification \mnilp\ is
consistent
with the
ghost number symmetry iff the generator $\epsilon$ is
assigned the
ghost number 2. If we compute the modification of the
action \rdin, 
we obtain  a better behaved integral
\eqn\mrdin{\int {{d\phi d X_{\alpha} d
\psi_{\alpha}}\over{{\rm
Vol}(G)}}
\exp \kappa_1 {\Tr} \left( i  \phi [ X_{1}, X_{2} ] -
{\half} \epsilon ( X^{2}_{1} + X_{2}^{2})
 +  \psi_{1} \psi_{2} \right).}
The factor $\kappa_1$ can be now reabsorbed into the $X$'s
and $\psi$'s
without
affecting the measure,  and  the $\psi$'s  can then be integrated
out. Also the matrices $X_{1}, X_{2}$ can be integrated out,
producing the
determinant
\eqn\mmrdin{ Z(N, \e, V)= \int {{d \phi}\over{{\rm Vol}(G)}}
{1\over{{\Det} \left(
ad (\phi) + \epsilon \right)}}.}

The supersymmetry $\delta$ allows a modification of the 
action  by the
observables
$$
S \to S + \sum_{n} T_{n} \CO_{n}
$$
where $\delta \CO_{n} = 0$.
In our case the operators $\CO_{n}$
are simply the gauge-invariant functions of $\phi$ as they are
also $U(1)_{\epsilon}$-invariant.
The simplest operators whose correlation functions may be
evaluated
are the gauge-invariant functionals of $\phi$, such as
${\Tr} (\phi^{l})$.

To summarize, we have shown that the computation of the
(regularized) correlation functions of the observables ${\Tr}
(\phi^{n})$
in the supersymmetric matrix integral \mit\ may be
reduced to the
computations of the integral over a single matrix $\phi$ of
the form
$V(z) =- \sum_{n} T_{n} z^{n}$:
\eqn\harvard{\int
{{d \phi}\over{{\rm Vol}(G)}} {{e^{- {\Tr} V(\phi)
}}\over{{\Det}
\left( ad (\phi) + \epsilon \right)}}.}
\noindent
{\it Remarks.} 
\item{1.} In ref.\index,  the similar
perturbation has been
used in the computations of the Witten index, which can
be reduced to
the
computation of the integral \mit\ for the group
$SU(N)/{\IZ}_{N}$
and without insertion of any observables. In that case
the result of
the computation was $\epsilon$-independent. Also, the
integral over
the
eigenvalues of the matrix $\phi$   was to be
understood
as a contour integral, to avoid the  contribution of the flat
directions that  corresponded to the unbound free particles. In our
case, the
flat directions contribute to the correlation functions
as well, and
the parameter $\epsilon$ serves as a regulator as in the
computations
of \higgs.

\item{2.} One may wonder about the physical meaning of the
$\e$-regularized integrals.
Here it is:
\eqn\ellgn{Z(N, \e, V)= {\Tr}_{\CH} (-)^{F} e^{-\b H}
e^{- \e \cdot J}
e^{-\Tr V(\phi)},}
where the trace is taken over the Hilbert space $\CH$ of
the quantum
mechanical system,
$H$ is the Hamiltonian, $F$ is the fermion number, $J$ is the
generator of the global
symmetry group (which we take to be $SO(d-2)$ for $d=4,6$
and $SO(6)$
for $d=10$,
see below). For example, in the case $d=4$,
 $J= 2i{\Tr} (\bar \phi [X_1,X_2])$.
 Just as in \sav,  this expression is
related to the
matrix
integral in the $\b \to 0$ limit. One can also consider  
directly the
quantum mechanical
path integral, i.e. the integral over the space of loops.  
In this
case the rational functions in
the formulas  \harvard\ and the similar formulas  below are 
replaced by
their trigonometric
counterparts. Also one can consider a $1+1$ model (Matrix  
strings) in which
case the ratio
of determinants leads to elliptic functions, just as in
\nikitathesis.

\subsec{Theory with eight supercharges}

This is the model obtained by the dimensional reduction of
$\CN=1$, $d=6$ theory.
In this model the index $\alpha$ of the matrices
$X_{\alpha}$,
$\psi_{\alpha} $ runs from $1$ to $4$. The symmetry
$U(1)_{\epsilon}$ is extended to $SO(4) \approx
SU(2)_{L} \times SU(2)_{R}$. The matrices $X_{\alpha}$,
$\psi_{\alpha}$
form two copies of the representation $(\half, \half)$ of
this group.
Also, the
fermion $\chi$ is promoted to a triplet $\vec\chi$, which is in
$(\half, 0)$.
The same metamorphosis is experienced by the auxiliary 
field $H \to
\vec H$.
The action is constructed by the same rules, the only
difference being that 
$$
\chi ( s - H) \to \vec \chi \cdot ( \vec s - \vec H),
$$
where
$$
s_{i} = [ X_{4}, X_{i} ] + {\half} \varepsilon_{ijk} [
X_{j}, X_{k}].
$$
The modification of the supercharge \mnilp\ is achieved by
introduction
of the generators $\epsilon_{L} \oplus \epsilon_{R} =
\left( {{\epsilon_{1} + \epsilon_{2}}\over{2}} \right)
\oplus \left(
{{\epsilon_{1} - \epsilon_{2}}\over{2}}
\right)$ of the Cartan
subalgebra of $SO(4)$.
The modified transformations are:
\eqn\msnilp{\eqalign{
\delta \psi_{1} = [ \phi, X_{1} ] + i \epsilon_{1} X_{2},
\quad & \quad
\delta \psi_{3} = [ \phi, X_{3}] + i \epsilon_{2} X_{4} \cr
\delta \psi_{2} = [ \phi, X_{2} ]  - i \epsilon_{1} X_{1},
\quad & \quad
\delta \psi_{4} = [ \phi, X_{4}] - i \epsilon_{2} X_{3} \cr
\delta \chi_{i} = H_{i} \quad & \quad \delta H_{3} = [\phi,
\chi_{3}]\cr
\delta H_{1} = [ \phi, \chi_{1}] + 2i \epsilon_{L}
\chi_{2} \quad
& \quad \delta H_{2} = [\phi, \chi_{2} ] - 2i \epsilon_{L}
\chi_{1}.\cr}}

Now we get, instead of  \harvard, the following
one-matrix integral
\eqn\harsb{\int
{{d \phi e^{ - {\Tr} V(\phi)}}\over{{\rm Vol}(G)}} {{
{\Det} \left( ad (\phi) + \epsilon_{1} + \epsilon_{2}
\right)}\over{
{\Det} \left( ad (\phi) + \epsilon_{1} \right)
{\Det} \left( ad (\phi) + \epsilon_{2} \right) }}.}

\subsec{Theory with sixteen supercharges}

It is of great interest to obtain the similar
expression for the integrals occurring in the reductions of
$d=10$ SYM.
In this theory the matrices $X_{\alpha}, \psi_{\alpha}$
have the index $\alpha$ transforming by the $\bf 8$ of the group
$SO(8)$. The antighost $\vec \chi$ belongs to $\bf 1
\oplus \bf 6$ of
$SU(4) \subset SO(8)$. Introduce the notation
$$
B_{i} = X_{2i-1} + i X_{2i}, i = 1,2,3,4.
$$
The matrices $B_{i}$ are in $\bf 4$ of $SU(4)$ and
$B_{i}^{\dagger}$
are in $\bf \bar 4$.
The ``gauge condition'' $\vec s$
splits as:
\eqn\ggcnd{\eqalign{\vec s = \mu \oplus \Phi \quad &\quad
\mu =
\sum _{i=1}^{4} [ B_{i} , B_{i}^{\dagger}] \in {\bf 1}\cr
\Phi_{ij} = [B_{i}, B_{j} ] + &
{\half} \varepsilon_{ijkl} [B_{k}^{\dagger},
B_{l}^{\dagger}],\cr
\Phi_{ij} = \epsilon_{ijkl} \Phi_{kl}^{\dagger}, \quad & \quad
{\rm i.e.}\quad  \Phi \in {\bf 6}.
\cr}}

The action constructed by the standard rules coincides
with that of the 
dimensional reduction of $d=10$, $\CN=1$ SYM. The gauge field
has ten components, which become $\phi, \bar\phi$ and
$X_{\alpha}$.
The sixteen-component fermion splits as $\psi_{\alpha}$, with
eight
components, $\vec \chi$ with seven components and $\eta$.

The global group $SU(4)$ (which is
not to be confused with the $R$-symmetry group of
$\CN=4$ SYM in four dimensions!)  as before allows us to modify the
supercharge $\delta$ in a manner analogous to \mnilp--\msnilp.
The Cartan   ganarator  of $SU(4)$ may be written as
$\epsilon_{1} \oplus \epsilon_{2} \oplus \epsilon_{3}
\oplus \left( \epsilon_{4} = - \epsilon_{1} - \epsilon_{2} -
\epsilon_{3}
\right)$.
The integrals \harvard--\harsb\ generalize to:
\eqn\harsh{\eqalign{\int &
{{d \phi e^{ - {\Tr} V(\phi)}}\over{{\rm Vol}(G)}} \cr
&  {{
{\Det} \left( ad (\phi) + \epsilon_{1}
 + \epsilon_{2} \right)
{\Det} \left( ad (\phi) + \epsilon_{2} + \epsilon_{3} \right)
{\Det} \left( ad (\phi) + \epsilon_{3} + \epsilon_{1}
\right)}\over{
{\Det} \left( ad (\phi) + \epsilon_{1} \right)
{\Det} \left( ad (\phi) + \epsilon_{2} \right)
{\Det} \left( ad (\phi) + \epsilon_{3} \right)
{\Det} \left( ad (\phi) + \epsilon_{4} \right) }}.\cr}}

\newsec{Determinant representation of the correlation
functions in the 
$d=4$ case}

In this section we study in detail the grand
partition function
$$
\CZ ( \mu, \e,  V) = \sum_{N=0}^{\infty} e^{\m N} Z(N,
\e,  V).
$$
 We show that $\CZ ( \m,\e,  V)$  has  a determinant
representation, very much like the correlation functions in
Sine-Gordon, and related models
are expressed in terms of Fredholm determinants \slavnov.

\subsec{Eigenvalue integral}

First we write the integral \harvard\ in terms of the
eigenvalues
$i\phi_1, ..., i\phi_N$ of the anti-Hermitian matrix $ \phi$:
\eqn\mint{Z(N,\e, V) = \int_{\IR^{N}} {{d\phi_{1}
\ldots
d\phi_{N}}\over{N! (2\pi \epsilon)^{N}}}
\prod_{i \neq j} {{\phi_i-\phi_j}\over{\phi_i-\phi_j + i
\epsilon}}
\prod_{i}
e^{-V({\phi_{i}})},}
where we changed  $V(i x)$ to $V(x)$.
The integral \mint\  can be rewritten, using the Cauchy formula, as
\eqn\mmint{Z(N, \epsilon, V) = \sum_{\sigma \in \CS_{N}}
(-)^{\sigma} \int{{d\phi_{1} \ldots d\phi_{N}}\over{N!
(2\pi i)^{N}}}
\prod_{k} {{e^{-V({\phi_{k}})}}\over{\phi_{k} -
\phi_{\sigma(k)} + i\e }  }.}

It turns out that the grand partition function (that is, with
 fixed chemical
potential) can be written as a Fredholm determinant of
an integral operator.
Let us
introduce the
notation
\eqn\ckle{W_{l}(\e, V) = \int_{\IR^{l}} \prod_{k=1}^{l}
{dx_{k}\over 2\pi } { e^{-V(x_{k})}\over{\e -i(x_{k} -
x_{k+1})}},
\quad x_{l+1} \equiv x_{1}.}
We may rewrite the sum over all the elements of the  permutation
group in
\mmint\
as the sum over the conjugacy classes that  are labelled by the
partitions
of $N$:
$$
N = \sum_{l=1}^{\infty} l d_{l}, \quad d_{l} \geq 0.
$$
Every permutation in the conjugacy class, labelled by
$\vec d = (d_{1},d_{2}, \ldots)$, is similar to the product
of cycles of lengths $1, 2, \ldots, $  the
number of  times the cycles with length $l$ appear
being precisely $d_{l}$.
The number of permutations in the given conjugacy class
$\vec d$
is equal to
$$
{{N!}\over{\prod_{l} l^{d_{l}} d_{l}!}}
$$
and the sign of any permutation in this class is
$(-)^{^{_{\sum_{l} d_{l}}}} (-)^{N}$.
Thus, \mmint\ may be represented as
\eqn\mmmint{Z( N, \epsilon, V) =  \sum_{\vec d: \sum
ld_{l} = N}
\prod_{l} {1\over{d_{l}!}} \left( - {{(-)^{l}W_{l}}\over{l}}
\right)^{d_{l}} }
and the grand partition function is equal to
\eqn\grptn{\CZ(\mu, \epsilon, V) = \exp  \sum_{l}
(-)^{l-1} e^{l\mu}\
{{W_{l}(\e, V)}\over{l}} .}
The quantity  $W_{l}(\e, V)$ may be represented as
 $W_{l} = {\Tr}
K^{l}$,
where $K$ is a   linear operator acting in the space of
functions
of one variable, as follows:
\eqn\opr{(K f)(x) = e^{-V(x)}
\int_{\IR} {dy\over 2\pi}
{{f(y)}\over{\e -  i(x - y) }} .}
Therefore, the grand partition function becomes
\eqn\grptfn{\CZ(\mu, \epsilon, V) = \exp \sum_{l}
{{(-)^{l-1}}\over{l}}
{\Tr} (e^{\mu} K)^{l}  = {\Det} \left( I + e^{\mu} K
\right). }

\subsec{Another representation for quadratic $V$}

Let us consider the case of a Gaussian potential
$V (x  ) = {1\over 2}
\left({\xi x\over \e}\right)^{2}$ and
 write  the partition function \mint\ again
as a matrix integral
\eqn\prtn{Z(N, \xi) = \int {{d\phi
dX}\over{{\rm
Vol}(G)}}
 \exp \left(
- {\half} {\Tr} [\phi, X]^{2} + {\xi^2\over \e^2} \Tr  
\phi^2  -
{\half} \epsilon^{2} {\Tr} X^{2} \right) . }
Considering  the matrices $X$ and $\phi$ as the
 Hermitian and anti-Hermitian part of the same complex  
matrix
$Z = (\e/\sqrt{ \xi}) X + (\sqrt{\xi}  / \e)
 \phi$, we rewrite  the matrix integral as
\eqn\rfrml{ Z(N,\xi)= {1\over{{\rm Vol}(G)}} \int dZ
dZ^{\dagger}
\ e^{-S}, \ \ \ \ S=  {\half} {\Tr} [ Z, Z^{\dagger} ]^{2} +
{{\xi}\over{2}} {\Tr} ZZ^{\dagger}. }
In polar coordinates
$$
Z = U H^{1/2}
$$
where $U$ is unitary and $H$ is a Hermitian matrix with
{\it positive}
eigenvalues $y_{1}, \ldots, y_{N}$, the measure and the
action read
$$dZ
dZ^{\dagger}= d U dH, \ \ \
S =  {\Tr} H^{2} - {\Tr} U^{-1} H U H + {\half} {\xi}  
{\Tr} H.$$
Using the  Harish-Chandra-Itzykson-Zuber formula  
\mehta\izii\ we
perform the $U$-integration  and find
\eqn\frmli{Z (N, \xi) = {1\over{N!}} \int_{\IR_{+}^{N}}
d^{N} y\
e^{- \sum_{i} \left( y_{i}^{2} + \half {\xi} y_{i}
\right) }
{\Det}_{ij}\left( e^{y_{i} y_{j}}\right).}
The grand partition function will be expressed
in terms of the
 quantities
\eqn\ccle{W_{l} = \int_{\IR_{+}^l} \prod_{i=1}^l d y_i
\ e^{ -\hf
 \left[  \xi y_i
+  (y_i - y_{i+1})^{2} \right]},  \quad\ \
(y_{l+1} \equiv
y_{1}),}
 as
\eqn\grnd{\CZ (\mu, \xi) = \exp
\sum_{l=1}^{\infty}(-)^{l-1} e^{\mu
l} {{W_{l}}\over{l}}. }
Clearly $W_{l} = {\Tr} \CK^{l}$, where
 $\CK$ now acts on the functions on the positive
semi-axis
as
\eqn\oprii{\CK f (y) = e^{-{{\xi}\over{2}} y}
\int_{0}^{\infty}
e^{-\half
 (y - y')^{2}} f(y') dy'. }
Therefore we arrive at the same determinant
representation \grptfn\
where the kernel $K$ is replaced by its Fourier transform
$\CK$.

For small $\xi$,
\eqn\asmt{(-)^{l-1} W_l =  {1\over{l \xi}} g_{l}({\xi}),}
where $g_{l}$ is an analytic function.

\newsec{The grand partition function as a tau-function}

In this section we show that  the grand partition function
\eqn\odve{\CZ(\mu , \e , V ) = \sum_{N=0}^{\infty} e^{\mu N} 
 \int_{-\infty}^{\infty} \prod _{i=1}^N dx_i \
e^{-V(x_i)}   \ \
{\prod_{i< j}
(x_i-x_j)^2\over
  \prod_{i, j}(x_i-x_j -i\e)}
   }
  can be represented as  a tau-function of the KP hierarchy.
 
\subsec{Vertex operator construction -- bosonic representation}

 Introduce  the  bosonic field
$\varphi(z) $ with mode expansion
   \eqn\henno{
  \varphi(z) = \hat q+ \hat p\ln z +\sum_{n\ne 0}
{J_{n}\over n}
z^{-n}, }
\eqn\frt{
[ J_n,  J_m]
=n\delta_{m+n, 0};   \ \  [\hat p,\hat q] = 1
}
and the vacuum state $|l\rangle$  defined by
\eqn\ljai{
J_n|l\rangle =0 , \ \  (n>0); \ \ \  \hat p
|l\rangle=l|l\rangle .
}
The associated normal ordering is defined by putting
$J_n$ with $n>0$ to
the
right.
Define the vertex operator
  \eqn\szso{
 \CV_\e (z) =:e^{ \varphi (z+i\e/2)}::
e^{ - \varphi(z-i\e/2)}:,
}
which satisfies
  the OPE
 \eqn\opevo{\CV_\e (z) \CV_\e (z')   = {(z-z') ^{2} \over
(z-z')^2
+\e^2}
:\CV_\e (z) \CV_\e (z')
: \ , }
the Hamiltonian
\eqn\hahaha{
H[t]=\sum_{n > 0}t_{n}J_n,
}
and the operator
\eqn\gegege{
\Omega_{\mu} =
\exp \Big(e^\mu   \int_{-\infty}^{\infty} {dz\over 2\pi
i} \CV_\e (z)
 \Big).
}
Then the  vacuum expectation value
  \eqn\pYiao{\tau_0[t]  = \langle 0|
 e^{ H[t]} \Omega _{\mu}|0\rangle
}
is equal to the canonical partition function \odve,
 with chemical potential $\mu$ and potential
\eqn\potKP{\eqalign{ V(z) = U \left( z + {i\over{2}} \e
\right) -U\left( z -
{i\over{2}} \e\right)& =-
\sum_{n=0}^{\infty} T_n z^n, \cr
U ( z) = - \sum_{n=1}^{\infty} t_n z^n, \cr}}
where  $\mu$ can be reabsorbed into the
definition of $V$
and
where the coefficients $T_n$ and $t_n$ are related by
\eqn\cpccS{\eqalign{
T_{n-1} = i \sum _{k=0}^{\infty} (-)^k
\left(^{n+2k}_{
n-1}\right){{\e^{2k+1}}\over{4^k}}
t_{n + 2k}.\cr}}

\subsec{Fermionic representation}

The fermionic representation of the partition function
is constructed  using the  bosonization formulas
  \eqn\bBo{\psi(z)= :e^{-\varphi(z)}:  \ \
 \psi^*(z)=
:e^{\varphi(z)}: \ \
  \partial\varphi(z)= :\psi^*(z)
\psi(z):,
 }
 where the
  fermion operators
 \eqn\pzpo{
        \psi(z)= \sum_{r\in \Z+ {1\over 2}}\psi_{r}
z^{-r- {1\over 2}}, \ \ \ \
        \psi^*(z)= \sum_{r\in \Z+ {1\over 2}}\psi^*_{-r}
z^{-r-
{1\over 2}}     }
satisfy the  anticommutation relations
 \eqn\cpmto{
        [\psi_{r},\psi^*_{s}]_+=\delta_{rs}.
         }
The  operators \hahaha\ and \gegege\ are represented by
 \eqn\curro{
        H[t] = \sum  _{n>0}t_n  \sum_r
:\psi^*_{r-n}\psi_{r}: \ \
(n\in \Z)
        }
 \eqn\araro{
        \Omega_{\mu}=\exp\Big[e^{\mu}
\int_{-\infty}^{\infty} {dx\over 2\pi i}
:\psi(x+i {\e\over 2} )\psi^*(x-i{\e\over 2}) :
        \Big].
        }
 and the
vacuum states with given electric charge $l$ satisfy
 \eqn\mnfio{\eqalign{
        \langle  l|     \psi_{-r} =\langle l| \psi^*_{r}&
= 0\ \ \ \
         \ (r>l )\cr
\psi_{r}| l \rangle =\psi^*_{-r}|l\rangle & = 0\ \ \ \
 \ (r> l). \cr}
        }
The original expression \odve\  is obtained
from  the expectation value  \pYiao\ by
first commuting the operator $e^H$ to the right
until it hits the right vacuum by
using the formulas
\eqn\evolU{\eqalign{ e^{H[t]} \psi(z)e^{-H[t]}& =
e^{\sum_{n=1}^{\infty}  t_{n}z^n}
\psi(z)\cr
e^{H[t]} \psi^*(z)e^{-H[t]}& = e^{-\sum_{n=1}^{\infty}
t_{n}z^n}
 \psi^*(z)\cr}
}
and then applying the Wick theorem
to calculate the expectation value
\eqn\fermP{\langle l |\prod
_{i} \psi(z_i) \psi^*(w_i)| l \rangle  =
\prod_{i} \left({z_i\over w_i}\right)^l \prod_{i<j} 
{(z_i-z_j)(w_i-w_j)\over (z_i-w_j)(w_i-z_j)}.
}


\subsec{The KP hierarchy}

The partition function  \pYiao\
is a particular case of the ``general solution'' of the KP
 hierarchy obtained as  the limit  $N\to \infty$ of a general
$N$-soliton solution \jimiwa
\eqn\TauKP{\tau_l[t] =  \langle l|e^{\sum_{n> 0}t_n J_{n} }
 \Omega_a|l\rangle,}
where the $ GL (\infty)$ rotation
\eqn\OmgG{  \Omega_a= \exp\Big( \int dx dy \ a(x,y)
\psi(x)\psi^*(y)\Big)}
is parametrized by
an  arbitrary integrable function $a(x,y)$\foot{In the soliton
solutions it  is a sum of delta-functions,
$a(x,y) = \sum_{k=1}^N  a_i \delta(x-p_i)\delta(y-q_i)$,
so that the
operator $\Omega_a$ is a finite sum,
$\Omega_a = \sum \sum_{k=1}^N  a_i\psi(p_i)\psi^*(q_i)
.$}.

The tau-functions $\tau_l, \ l\in \Z$, are Fredholm
determinants
\eqn\frtt{
\tau_{l}= \det (1+ e^\mu  K_l )
 }
of the kernels $ K_l$
\eqn\KerNo{ K_l( x, y)= {E_l(x+i\e)\over
E_l(x-i\e)}\ \ {1\over
x-y -2i\e},
\ \ \ \ \ \
E_l(x)=  x^l\  \exp\left(\sum_n t_n x^n\right).}
For $l=0$ we get precisely the operator $K$ \opr.

The KP  hierarchy of differential equations
 is generated by
 the {\it Hirota bilinear equations} \hirota :
\eqn\Hiilr{
\oint{dz\over 2\pi i} z^{l-l'}
\exp\Big(\sum_{n>0} (t_n-t'_n) z^n\Big)
\tau_{l}\left(t_n-{1\over n} z^{-n }\right)
\ \tau_{l'}\left(t'_n+{1\over n} z^{-n }\right) =0\ \ \ \ \ (l'\le l).}

Let us sketch the proof of \Hiilr .
  First we remark that  each
element $\Omega\in
GL(\infty)$ is represented by an infinite c-number matrix
$a = \{
a_{rs}\}$
\eqn\fermA{ \Omega \psi _r \Omega ^{-1}
= \sum_s \psi_s a_{sr}, \ \ \ \
\Omega ^{-1}\psi _r^* \Omega
= \sum_s  a_{rs}\psi_s^*.}
As a consequence, there exists a  tensor Casimir operator
\eqn\tenC{S_{12}= \sum _{r} \psi_r\otimes
\psi^*_r= \oint
{dz\over 2\pi i} \psi(z)\otimes  \psi^*(z)}
which satisfies, for any   $\Omega\in GL(\infty)    $,
\eqn\tenCc{S_{1 2} \ \Omega \otimes  \Omega =  \Omega
\otimes  \Omega
\ S_{1 2}.}
On the other hand  $S_{12} \ |l\rangle
\otimes  |l\rangle  =0$  because, according to \mnfio,
 for each $r$ either $\psi_r$ or $\psi^*_r$
is annihilated by the right vacuum $|l\rangle$.
Therefore  \tenCc\ implies that $S_{1 2} \ \Omega
|l\rangle \otimes
\Omega|l\rangle  = 0.$
  Taking the scalar product with
$\langle l +1 \vert  e^{H[t]}\otimes \langle l -1\vert
e^{H[t']}$ we
find
\eqn\hirOp{
\oint {{dz}\over{2\pi i}}
\langle l+1 \vert  e^{H[t] } \psi (z) \Omega_{\mu} \vert l
\rangle
\langle l-1  \vert  e^{H[t'] } \psi^{*} (z) \Omega_{\mu}
\vert l \rangle
= 0,
}
where the integration contour surrounds the origin.
Equation  \hirOp\ simply reflects the fact that the tensor Casimir
\tenC\ is 
constant on the orbits of $GL(\infty)$.
 Finally we use the bosonization formulas \bBo\
to  represent
the fermions as vertex operators,
$\psi(z)\to{\bf V} _-(z)$, $\psi^*(z)\to {\bf V}_+(z)$,
where ${\bf V}_{\pm}(z)$ act in the space of the coupling
constants as
 \eqn\vop{
{\bf V}_{\pm}(z)= \exp\Big(
 \pm \sum_{n=0}^\infty t_n
z^n \Big)
\exp\Big( \mp \ln {1\over z} \ {\partial\over \partial \mu}
\mp
\sum_{n=1}^\infty
{ z^{-n}\over n}
 {\partial\over \partial t_n} \Big).
   }
The general case
  $l\ne l'$ is treated similarly, and one obtains  the
following
identity
\eqn\hireqn{
\oint  dz \   \Big({\bf V}_+(z)\cdot \tau_l[t]
\Big)\
\Big({\bf V}_-(z) \cdot \tau_{l'}[t']\Big)
=0  \ \ \ \ \ \ (l'\le l ), 
   }
which is identical to  the
Hirota equation  \Hiilr .

The differential equations of the KP hierarchy are obtained by
expanding \Hiilr\ 
 in the differences $y_n= \hf(t_n - t'_n)$.
In  the case  $l'=l$, the first non-trivial equation
(the KP equation) is obtained by
requiring that   $y_1^3$ term vanishes:
\eqn\olki{
\Big({\partial^4\over\partial y_1^4}+3{\partial^2\over\partial
y_2^2}-4
{\partial\over\partial y_1}{\partial\over\partial
y_3}\Big)\tau_l[t+y]
 \tau_l[t-y]\Big|_{y=0}=0.
  }
In terms of  the ``specific heat''
\eqn\spheat{u[t]= 2
{\partial^2\over\partial t_1^2} \log \tau_l,} 
the KP equation
reads
\eqn\KPe{
3{\partial^2 u\over\partial t^2_2} +
{\partial\over\partial t_1}
\Big[ -4 {\partial u\over\partial t_3} +6 u {\partial
u\over\partial
t_1}
+{\partial^3 u\over\partial t_1^3}\Big]=0.
  }

\newsec{ The $d=4$ integral with  quadratic potential: KP
equation, weak coupling and double scaling}

In this section we study
  the case $t_n=0, \ n>3$.

\subsec{Reduction to a  single  equation}
 
A potential of the form  $V({x} ) = -\mu + \lambda
x^2$, $\lambda
= \xi^{2}$
is related to  the
three couplings $t_1,t_2, t_3$ by
\eqn\muxi{\mu =  i \e \left( t_1- {1\over{4}} \e^2 t_3 -
{t_2^2\over
3t_3} \right), \ \
\xi^{2}  = - 3i\e t_3.}
Rescale $\e \to 1$. Then $u = - 2 \p_{\m}^2 \CF$ and it
is easy to
show that \KPe\ implies the following partial differential
equation for the function  $\psi = \xi u(\mu, \xi)$:
\eqn\hpf{\psi_{\xi} + \psi \psi_{\mu} + {{\xi}\over{6}}
\left( \psi -
\psi_{\mu\mu} \right)_{\mu}
= a(\xi).}
 By comparing it with  the expansion in
\grnd--\asmt, 
we see that $a({\xi}) \equiv 0$. If we expand
\eqn\ans{\psi (\mu, \xi) = \sum_{l=1}^{\infty} e^{\mu l} e^{-
{{{\xi}^{2}}\over{12}} ( l^{3} - l) } \eta_{l}({\xi}),}
then eq.\hpf\ is equivalent to the infinite system
of recursive
first-order differential equations:
\eqn\cpld{\eta_{l}^{\prime} = {{l}\over{2}} \sum_{p+q =
l} \eta_{p}
\eta_{q}
e^{-{\xi^{2}\over{4}} l pq}.}

The form \ans\ is dictated by the semi-classical
approximation to the
integral \mrdin. Indeed,
the expression ${{{\xi}^{2}}\over{12}} ( l^{3} - l)$ is
nothing but
the classical action evaluated on the solution to the
equations of
motion:
\eqn\sdlp{\eqalign{ [X_1, X_2] & = - 2i \l \phi, \cr
[\phi, X_2] & = - i X_1 , \cr
[\phi, X_1] & = + i X_2. \cr}}
The solutions to \sdlp\ are classified by the
decompositions of
$N$-dimensional
representation into irreducibles of $SU(2)$. The
logarithm of the
grand partition function
takes into account only irreducible
$l$-dimensional
representations,
the rest is generated by the exponentiation. The
functions $\eta_l$
 therefore describe
the  quantum   fluctuations around the saddle points.

We conclude this section
by listing the two equivalent forms of the equation
obeyed by $u$:
\eqn\KPee{\eqalign{& \psi_{\xi} + \psi \psi_{\mu} +
{{\xi}\over{6}} ( \psi - \psi_{\m\m})_{\m} = 0\cr
& 2 u_{\l} + {1\over \l} u +  u u_{\m} + {1\over 6} (u -
u_{\m\m})_{\m} = 0.\cr}}

\subsec{Weak coupling limit}
For low values of $l$, eqs. \cpld\ can be solved
explicitly.
It is interesting to look at the large $\lambda = \xi^2$
asymptotics of the
solutions.
We expect that
as $\lambda \to \infty$ the partition function at fixed
$N$ has the
following scaling behaviour of $Z(N, \l)=e^{F(N,\l)} $:
\eqn\scal{
Z(N, \l) = {1\over{\lambda^{N^{2}/2}}} \left( 1 +
\ldots \right) =
{1\over{\xi^{N^{2}}}} \left( 1 + \ldots \right).
}
On the other hand it follows  from \cpld\ that 
\eqn\low{u = - {{e^\m}\over{\xi}}  + {{e^{2\m}}\over{\xi^2}}
\sum_{n=0}^{\infty} {{(2n-1)!!}\over{\xi^{2n}}} + \ldots }
and therefore indeed
\eqn\lowu{Z(1, \l)  = {1\over{2\xi}}, \quad
Z(2, \l)  = - {1\over{\xi^{4}}}
\sum_{n=1}^{\infty} {{(2n-1)!!}\over{\xi^{2n}}},}
in accordance  with the scaling \scal .

It turns out that equation \hpf\ has another interesting
property.
Suppose we are studying the  't Hooft limit,
where the free energy has an expansion
of the form
\eqn\FNmu{
F(N, \lambda) = \sum_{g=0}^{\infty} N^{2 - 2g}
F_g({{N}/\lambda}).
}
Given $\CZ(\mu, \lambda) = e^{\CF ({\mu}, \lambda)}$ we
extract the
fixed-$N$ partition
function via the Fourier transform
\eqn\frtr{Z(N, \lambda) = \oint {{d \mu}\over{2\pi i}}
e^{-i \mu N +  \CF(i \mu, \lambda)}  ,}
which can be taken, in the large-$N$ limit,  using
the saddle-point approximation. In the planar limit ($N
= \infty$,
with $\l/N$ finite) the functions
$\CF (\mu, \lambda)$ and $F (N,\lambda)$ are Legendre
transforms of each other. If we keep all the $1/N$
corrections,  then
the corresponding
expansion of $\CF(\mu, \lambda)$ is:
\eqn\expns{\CF (\mu, \lambda) = \sum_{g=0}^{\infty}
\lambda^{2 -2g}\CF_g
\left({{\mu}/{\lambda}}\right)  }
and
\eqn\expnsii{u (\mu, \lambda) = \sum_{g=0}^{\infty} u_g
\left({{\mu}/{\lambda}}\right) \lambda^{-2g}, \qquad
u_g(x) = -
2 \CF_g^{\prime\prime} (x).}

Let us introduce the  variables
\eqn\xychi{x = {{\mu}/{\lambda}},\  \ y= {\lambda}^{-2} ,\ \
\chi (x,y) = u(\mu, \lambda),}
which are relevant for the 't Hooft limit.
Then eq. \hpf\ may be rewritten as:
\eqn\hpff{\chi - 2\left( x - {1\over{12}}\right)
\chi_{x} \ - 4y \chi_{y} + \chi \chi_{x} - {{y}\over{6}}
\chi_{xxx} =
0,}
which after the expansion
\eqn\thftexpn{\chi (x, y) = \sum_{g=0}^{\infty} y^{g}
\chi_g(x)}
reduces to the infinite system of  recursive equations for
$\chi_g$'s:
\eqn\rcrs{\eqalign{&
\left(  {1\over{6}} - 2x  + \chi_{0} \right)
\chi_{g}^{\prime}
+ \left( 1 - 4g - \chi_{0}^{\prime} \right) = 
{{\chi_{g-1}^{\prime\prime\prime}}\over{6}} - \sum_{a+b
=g, a,b \neq
0} \chi_{a} \chi_{b}^{\prime}, \quad g > 0 \cr
& \qquad \qquad \chi_{0} - 2 \left( x - {1\over{12}} \right)
\chi_{0}^{\prime} + \chi_{0} \chi_{0}^{\prime}  = 0.\cr}}
The equation for $\chi_{0}$ is the only non-linear one.
Its solution is:
\eqn\sltn{ x = {{\a_{0}}\over{2}} \chi_{0}^{2} + \chi_{0} +
{1\over{12}}. }

Of course eq. \hpf\ has more general solutions, in
particular
those for which the expansion \thftexpn\ is not bounded
as $g \geq
0$.
It turns out that the solution corresponding to the
matrix integral
in question does have the form \thftexpn.
In Appendix  A  we show that $\a_{0} = - {\pi^2}$. 

It follows  that, for large $N$ and $\lambda$,
the free energy has 't Hooft-like behaviour\foot{Recall
that $F = \CF
- \m N$ in our conventions.}:
\eqn\asm{F(N, \lambda) = - {1\over{10}} \left( {{ 243
\pi^{2}}\over{4}} \right)^{1/3}  N^{2} \left(
{{\lambda}\over{N}}
\right)^{1 \over 3} + \ldots }

\subsec{Double scaling limit near the quadratic
singularity in
$\chi(x)$}

So far we investigated  \mint\ only in the large-$N$ ('t
Hooft) limit
in the canonical ensemble or,  equivalently, in the
 large-$\mu$ limit for
the grand canonical ensemble.

As we know from \dsl\ and  Appendix B, the
universal scaling behaviour of higher $1/N$ corrections
sometimes can
 be summed up to some functions obeying non-linear differential
equations, such as  Painlev\'e $\II$ for the pure $2d$ gravity.

It is reasonable to ask whether we can do the same with
the $1/\mu$
expansion for our model
starting from the general KP equation \hpff \ and what  the physical or 
geometrical meaning of 
 this expansion is (we recall that in the pure gravity described by the one-matrix model of Appendix  B the corresponding $1/N$ expansion has 
the meaning of the  expansion over the genera of the topologies of the two-dimensional manifold).

Let us  concentrate on the square-root singularity
of \sltn\ at
$\chi_{c}={1\over \a_0}$, $x_{c}= - {1\over 2 \a_0} +{1\over 12}$.
We try the following ansatz:
\eqn\ansz{\chi=\chi_{c} + y^a {\up} ( z ), \quad  z = y^b
\left(x-x_{c}\right) .}
As in the case of the one-matrix model,
 the presence of a quadratic singularity implies that $b =
- 2a$.
In  full analogy with   Appendix B (the  only
difference being that
$y$ takes the place of $1/N^2$),
inserting  this ansatz into \hpff,  and neglecting the
subleading
corrections, we obtain
$b =-{2\over 5}$, $a={1\over 5}$ and that function
$\up (z)$
satisfies the Painlev\'e $\II$ equation:
\eqn\pain{\up^{\prime\prime}-3\up^2+
{{12}\over{\pi^2}} z =0.  }

{}From this equation we find the following coefficients of the $1/\lambda$ 
expansion in \expns\ (which is the same as the $1/\mu$ 
expansion in this approximation)  for the singular part of the free 
energy near the critical point:
\eqn\frexm{\CF_0^{\rm sing}=-\Delta ^{5\over 2}, \ 
\CF_1^{\rm sing}= - {1 \over 24} \log 
\Delta, \ \cdots , }
where $\Delta={\rm const} (x-x_c)$ (we choose the constant  in such a way that the 
coefficient in front of 
$\Delta ^{5/2}$ be $-1$, then the next coefficients, $-{1\over 24}, \cdots$,   are 
universal constants).

So everything goes just as  in the pure  $2d$
quantum gravity. The $1/\mu$ expansion looks like
the topological $1/N$ expansion, the coefficients
giving the leading scaling behaviour of the
partition functions of successive topologies (see the details in \dsl
). It is tempting to speculate
that the quadratic singularity in $\chi(x)$ corresponds
to the pure gravity. It may be
related to the large planar graph expansion with
respect to $g$ in the model of dense self-avoiding random  paths  (we will 
argue at the end of 
section $6$  that our matrix integral describes  such a model in the large-$N$ limit).
It would be interesting to demonstrate it by passing from the grand
canonical to the canonical ensemble for the free energy.

\newsec{Saddle-point approach}

\subsec{$d=4$ integral}

So far
we managed to calculate the grand canonical version of
the integral
 \grptn\ in the large-$\mu$ limit by the use of the KP
equations.
It is not clear whether we
can derive  from this asymptotics the large $N$ limit of canonical partition
function. In
fact, we shall show that it is possible
by comparing the results  with a  more direct approach,
originally
proposed in
this context by Hoppe \hoppe.
Namely, in the case of Gaussian potential $V(x)=
\l x^2$, it
is possible to
 solve the integral saddle-point equation for the
distribution of the eigenvalues of a
matrix  in \mint.
We work out the details of the solution (correcting some minor
mistakes in \hoppe\ and actually deriving the result) and
extract interesting critical behaviours of our system.

It is natural to
scale the coupling $\lambda$ as $N$, and rescale $\e$ to
$1$, i.e.  to
set:
\eqn\thoo{\l = {{N}\over{g^{2}}}, \quad \e \to 1.}
Indeed by rewriting the integral \mint\ as:
\eqn\mintii{Z (N, \lambda) = {1\over{{\rm vol} (G)
\l^{{N^{2}}\over{2}}}} \int
dX d\phi \exp \left( -{\Tr} \left( \phi^{2} + X^{2}
\right) +
{1\over{\lambda}} {\Tr} [ X, \phi]^{2} \right)} 
we see that ${1\over{\sqrt{\l}}}$ plays the role of the
coupling
constant while
${N\over{\l}}$ is the 't Hooft coupling. 
In the large-$N$ limit the integral \mint\ localizes onto
the critical point of the effective potential:
\eqn\effpot{V_{\rm eff}(\phi) = \sum_{i} V(\phi_{i}) +
\sum_{i < j}
{\rm log} \left( 1 + {1\over{(\phi_{i} - \phi_{j})^{2}}}
\right).}
Its critical point is found from the equation:
\eqn\spec{{{\phi_k}\over{g^{2}}} ={1\over{N}}
\sum_{j\ne k}{1\over  \left(\phi_k-\phi_j\right) \left(
1+(\phi_k-\phi_j)^2\right)}.}
In the usual fashion we assume that the
eigenvalues in the large-$N$
limit form a continuous medium of  density
\eqn\dnsty{\rho (x) = {1\over{N}} \sum_{i} \delta ( x -
\phi_{i}).}
When $g$ is real, it is natural to  expect  
 $\rho$ to vanish outside of the interval
$[-a, + a]$,
and to be an even function $\rho(x) = \rho(-x)$. We
introduce the resolvent:
\eqn\fyfy{W(z) = \int_{-a}^{+a} dx {{\rho(x)}\over{z-x}},}
and rewrite the eq. \spec\ as:
\eqn\spee{{{2x}\over{g^{2}}} =  W(x+i0) +  W(x-i0 )- 
 W(x+i) - 
W(x-i),}and 
where it is assumed that $x\in [-a, a]$.
 The spectral density is given by the discontinuity of 
$W(z)$ along the cut at $(-a, +a)$: 
\eqn\fhu{W(x+i0) - W(x- i0) =- 2\pi i \rho(x).}

Let us introduce the  holomorphic function
\eqn\derv{
G(z) = {{z^{2}}\over{  g^{2}}} + i \left[W\left(z +
{i\over{2}}\right) 
- W\left( z -
{i\over{2}}\right) \right]}
which is related  to the 
expectation value of the fermionic current
 $\langle \psi^*(z)\psi(z)\rangle$ in the grand 
canonical ensemble \odve .
  The saddle-point equation \spee\ can now be rewritten as
\eqn\sddle{G\left(x + {i\over{2}}\right) = G\left(x - {i\over{2}}\right), \quad
x \in ( -a,
+ a).}
The definitions  \fyfy \ and \derv\ imply that, in case of positive 
coupling  $\l = N/g^2$
\eqn\prop{\eqalign{W(z) = {\overline{W({\zb})}}, \quad &
W(z) = - W(
-z),\cr
G(z) = {\overline{G({\zb})}}, \quad & G(z) = G(-z), \cr}}
and also that the function $G(z)$ has the cuts at $(\pm
{i\over{2}} -
a,
\pm {i\over{2}} +a)$. It is also clear from \sddle \ and \prop\
that $G(z)$ is real when $z \in \IR, i\IR, (\pm
{i\over{2}} - a,
\pm {i\over{2}} +a)$. Hence $G(z)$ defines a holomorphic
map of the region $\CU$ bounded by $\IR_{+}, i\IR_{+}$
and by the
sides
of the interval $({i\over{2}}, {i\over{2}} + a)$ onto
the upper half-plane $\CH$. The inverse map is given by
the following
integral formula: $G(z) = \zeta$, where:
\eqn\mapp{z=A\int_{x_{1}}^{\z}
{dt (t-x_{3})\over \sqrt{(t- x_{1})(t-x_{2})(t-x_{4})}}.    }
The map acts on  the special
  points $x_{1} > x_{2} > x_{3} > x_{4}$
 and $\infty$
as follows:
\vskip 11pt
\hbox{\qquad\qquad\qquad\qquad\qquad\qquad
\vbox{\offinterlineskip
\hrule
\halign{&\vrule#&\strut\quad\hfil#\quad\cr
height2pt&\omit&&\omit&\cr
& $\z$ \hfil && $z$ &\cr
height2pt &\omit &&\omit&\cr
\noalign{\hrule}
height2pt&\omit&&\omit&\cr
& $+\infty$ && $+\infty$ &\cr
& $x_{1}$ && $0$ &\cr
& $x_{2} $ && ${i\over{2}}$ &\cr
& $x_{3}$ && ${i\over{2}} + 
a$ &\cr
& $x_{4}$ && ${i\over{2}}$ &\cr
& $-\infty$ &&$+ i\infty$ &\cr
height2pt&\omit&&\omit&\cr}\hrule}
}
These conditions imply the following equations on $x_{i},
A, a$:
\eqn\cndti{\eqalign{\half = A & \int_{x_{2}}^{x_{1}} dt {{t -
x_{3}}\over{\sqrt{
(t - x_{2})(x_{1} - t) (t - x_{4})}}} \cr
a = A & \int_{x_{3}}^{x_{2}} dt {{x_{3}- t}\over{\sqrt{
(x_{2} - t)(x_{1} - t) (t - x_{4})}}} \cr
a = A & \int_{x_{4}}^{x_{3}} dt {{x_{3}- t}\over{\sqrt{
( x_{2}- t )(x_{1} - t) (t - x_{4})}}}. \cr}}
{}From \derv\ we know the large-$z$ asymptotics of $\zeta$:
\eqn\lrgz{\eqalign{\zeta &=  \quad  {{1}\over{g^{2}}}  z^{2}+
z^{-2} + \delta
z^{-4}+ \ldots , \cr 
z &=  \quad  g \z^{\half} - {1\over{2g}} \z^{-{3\over{2}}}  -
{{\delta}\over{2 g^{3}}} \z^{-{5\over{2}}} +  \ldots,\cr}}
where
\eqn\dlt{\delta = - {1\over{4}} + 3 \nu, \quad \nu =
\int_{-a}^{+a}
\rho(y) y^{2} dy. }

At large $t$ the function $z(\z)$, as given by \mapp,  has the
following expansion:
\eqn\expa{z= 2 A \left( {\z}^{\half} + a_{0} + a_{1}
{\z}^{-{1\over{2}}} + a_{2} {\z}^{-{3\over{2}}} + a_{3}
{\z}^{-{5\over{2}}} + \ldots
\right),}
where
\eqn\coff{\eqalign{a_{0} = &
\int_{x_1}^{\infty} dt \left(  { (t-x_{3})\over \sqrt{(t-
x_{1})(t-x_{2})(t-x_{4})}}  - {1\over{\sqrt{t - x_1}}}
\right) \cr
a_{k } = &  {{(-)^{k-1} }\over{2k-1}} \left( \g_{k} +
x_{3} \g_{k-1}
\right) , \quad k > 0 \cr
\g_{k}  = & \sum_{ p + q  + r = k}  \pmatrix{-\half \cr p}
\pmatrix{-\half \cr q}\pmatrix{-\half \cr r} x_{1}^{p}
x_{2}^{q}
x_{4}^{r} .\cr}}
By comparing \lrgz\ and \coff, we get the following
equations on
$x_{i},  A$:
\eqn\cndt{\eqalign{A \quad &\quad =  \quad{{g}\over{2}} \cr
a_{0} (x_{i}) \quad & \quad = \quad 0 \cr
 x_{1} + x_{2} +  x_{4} & \quad = \quad 2 x_{3} \cr
  x_{1}^{2} + x_{2}^{2} + x_{4}^{2}  - 2x_{3}^{2} &\quad
=  \quad
{6\over{ g^{2} }} , \cr}}
which, together with eqs. \cndti,  fixes everything
completely. As shown in \hoppe,  the equation $a_{0} = 0$
follows from
 \cndti\ by a contour-deformation argument. We introduce
more notation:
\eqn\lstnt{\eqalign{y_{i} = g x_{i}, \quad & \quad
\lambda_{i} =
{y_i \over  y_{1} - y_{4}}  \cr
m= {{y_{2} - y_{4} }\over{y_{1} - y_{4}}} , \quad  1 >  &
m > 0,
\quad m^{\prime} = 1- {1\over{m}}, \cr}}
with which eqs. \cndti\ assume the following form:
\eqn\trans{\eqalign{( y_2 + y_4 - y_1) {\bf K} (m) + & 2
( y_1 - y_4)
{\bf E} (m)  = 0 \cr
- ( y_1 + y_2 - y_4) {\bf K} (m^{\prime}) + & 2 ( y_2 -
y_4) {\bf E}
(m^{\prime})  = \sqrt{{y_{2}- y_{4} }\over{g}} ,\cr
}}
where we use the standard elliptic functions
\eqn\ellptc{{\bf K}(m) =  \int_{0}^{\pi \over{2}}
{{d\theta}\over{\sqrt{1 - m {\rm sin}^{2}\theta}}}, \quad
{\bf E}(m) =  \int_{0}^{\pi \over{2}}  d\theta \sqrt{1 -
m {\rm
sin}^{2}\theta},}
which have the following crucial properties:
\eqn\ellptcp{\eqalign{{\bf K}(m^{\prime})  = \sqrt{m}
{\bf K} ( 1-
m), \quad & \quad
{\bf E}(m^{\prime}) = {1\over{\sqrt{m}}} {\bf E}(1 - m) \cr
{\bf E} (m) {\bf K} ( 1- m) + {\bf E}(1- m)&  {\bf K}(m)
-  {\bf
K}(m) {\bf K}(1- m) = {{\pi}\over{2}} .\cr}}
In the sequel we use the short-hand notations  ${\bf E}
= {\bf E}(m), {\bf K} = {\bf K}(m), \vartheta = {\bf
E}/{\bf K}$.
The first equation in \trans\ allows us to express $\l_2$
in terms of
$m$, while the second,
together with \ellptcp, gives $g (y_1 - y_4)$:
\eqn\tofm{\l_2 \ =\quad1 - 2 \vartheta,\quad g (y_1 -
y_4)   = {1\over{\pi^2}} {\bf K}^2.}
{}From \cndt\ and  the equations $\l_4 = \l_2 - m$ and $  \l_1 = \l_4+1$
we get  an expression for $y_{1}- y_4$ and
consequently for $g$:
\eqn\yofm{\eqalign{& g ^2(m) =  {{\bf K}^4 \over 3 \pi^4 }
\left( 4 m \l_2 + 1 - 3\l_2^2 - 2\l_2 \right) \cr
& = {{\bf K}^4 \over 3 \pi^4 }
  \left( - 3 {\vartheta}^2  +  2
(2 - m) {\vartheta} - ( 1- m)  \right).\cr}}
We can now compute ${1\over N} \langle {\Tr} \phi^{2} \rangle = \nu$ and
$F(N, g)$:
\eqn\onpnt{\eqalign{ &\nu (m)  =   \quad
{{g^4}\over{N^{2}}} {{\p F(N, g)}\over{\p g^2}} \cr  &=   
  {1\over{12}}+ {{{\bf K}^2}\over{5 \pi^2}} \left(
{{4 m \l_2 (
1- m) + (5 \l_2^2 -1 )  ( 2 m  - 1 - \l_2)  }\over{4 m
\l_2 + 1 -
3\l_2^2 - 2\l_2}} \right)\cr
 &=   {1\over{12}} - { {{{\bf K}^2}  
\over{5\pi^2}}}{ { {10\vartheta}^2({\vartheta}+m-2)+2\vartheta}(6-6m+m^2)+(1-m)(m-2)
\over    3 {\vartheta}^2 +  2 (  m -2 )
{\vartheta}  +  1- m  } . \cr}}
The formulae   \yofm \ and \onpnt\  give  the exact analytic
solution of
the
large-$N$ model in the parametric form.

{\it  Small-$g$ expansion}. In this case we expect to get
a regular
planar graph expansion of the matrix
integral \prtn\ with respect to the quartic term in the
action. The
careful analysis shows that the $g \to 0$ limit
corresponds to $m \to 0$. In this limit we can expand:
\eqn\gandnu{\eqalign{g^2 = & {1\over{128}} m^2  +
{1\over{128}} m^3 + {{119}\over{16384}} m^4
+ \ldots \cr
\nu = & {1\over{256}} m^2 + {1\over{256}} m^3 + 
{133 \over 4096\sqrt{2}} m^4+ \ldots , \cr}}
which implies
\eqn\nuofg{\eqalign{\nu = & {1\over{2}} g^2 - {1\over{2}}
g^4 +
\ldots \cr
F (N, g) = &  N^{2} \left( {1\over 2} {\rm log}g^2 -  {1\over 2} g^2
+ \ldots
\right) \cr}}
in perfect agreement with the planar graph expansion
and 
formula \scal.

{\it Large-$g$ limit}. This limit  corresponds to the situation
where we are
far beyond the convergence
radius of the $g$-series. The integral \prtn\ is
dominated by the
commutator term in the action
and hence the fluctuations of the matrices are very
large due to the presence of the zero modes.
It follows from the careful study of \tofm\ and \yofm\ that the $g
\to \infty$
limit
corresponds to $m \to 1$. In this limit ($m = 1 - \ve$):
\eqn\expnss{\eqalign{g^2 = &  {1\over{12\pi^4}} {\rm
log}^{3}\left( {16 \over \ve}\right) -
{{1}\over{4\pi^4}} {\rm log}^{2} \left( {16 \over
\ve}\right)\ldots \cr
\nu  = &  +{1\over{20\pi^{2}}} {{\rm log}^2 \left( {16
\over \ve}
\right)}  - {7\over{20 \pi^2}}  {{\rm log} \left( {16
\over \ve}
\right)}  + \ldots  \cr}}
Hence
\eqn\rhoofg{\eqalign{\nu = &  {{(12\pi)^{2\over
3}}\over{20}}
g^{4\over 3} - {3\over{(12 \pi)^{2\over 3}}} g^{2\over
3}\ldots\cr
F(N, g) = &  - N^{2} \left(  {{3 (12\pi)^{2\over
3}}\over{40}} g^{-{2\over 3}}  - {9\over{5(12 \pi)^{2\over
3}}}    g^{-{5\over 3}} + \ldots \right),\cr}}
in perfect agreement with \asm !\foot{Indeed,
${1\over{10}} \left(
{{243 \pi^{2}}\over{4}} \right)^{1\over 3} = {{3
(12\pi)^{2\over
3}}\over{40}} $.}.   In fact, the strong coupling
expansion can be
greatly simplified if we choose $L = {1\over{{\rm log}
\left( {16
\over \ve}\right)}}$ as an expansion parameter and
systematically
neglect all
non-perturbative (in $L$) corrections, i.e. we consider
the leading
logarithmic approximation. We then get  the very simple formulas :
\eqn\ledlog{\eqalign{g^2 = & \ \ {1\over{12 \pi^4 L^3}}
\left( 1 - 3
L\right)\cr
\nu = & \ \  {1\over{20 \pi^2 L^2}} \left( {{1 - 10L + 20
L^2}\over{1 - 3 L}} \right) + {{1}\over{12}} .\cr}}
The surprise is that  formulas \ledlog\ are {\it
exactly} equivalent
to \sltn.

\noindent
{\it Proof.} \ Equation \sltn\ with $\a_0 = {\pi^2 \over 2}$ gives:
\eqn\drfr{\CF_{0}^{\prime} = {\pi^2 \over 12} \chi_{0}^3
- {1\over
4} {\chi_{0}^2} , \quad \CF_{0} =
{\pi^4 \over 120} {\chi_{0}}^5 - {5\pi^4 \over 96}
{\chi_{0}}^4 + {1\over
12} {\chi_{0}}^3.} The
Legendre
 transform leading from $\CF_{0}$ to
$F_{0}$ yields: $\nu = x - 2 {\CF_{0} \over
\CF_{0}^{\prime}}$, while
$g^2 = - \CF_{0}^{\prime}$. Clearly, it leads to \ledlog\
if we
substitute
\eqn\logu{{\chi_{0}} = {{{\rm log} \left( {16 \over
\ve}\right)}\over{\pi^2}}.}
This confirms once again that indeed $\a_0 = {\pi^2 \over 2}$.

\noindent
{\it Remarks.} \item{1.}It is very tempting to speculate
that the
relation to supersymmetric gauge theories,
which was one of the original motivations of this work, is
somehow
revealed by the appearance
of a family of elliptic curves, parametrized by the value of
coupling $g$
just as in \SeWi. Notice that the solution that we
studied here
has something to do with mirror symmetry. Indeed,
the naive coordinates in the space of our Lagrangians
(which is just
the coupling $g$ for a
quadratic potential) have been replaced by the period of a
certain
differential on the elliptic curve.
It is conceivable that a similar construction takes
place for more
general potentials.

\item{2.} 
The partition function does not change, in the large-$N$ limit, if we
substitute the commutator by the anticommutator in the action in
\prtn. The latter model generates the statistical ensemble of
$\phi^4$-type random graphs covered by dense non-oriented
self-avoiding random loops and has been studied in \kristjansen .  It
describes the dense phase of the $O(n)$ loop-gas model \kostovOn\ with
$n=1$.  It also can be viewed as a matrix model counting 3-colored
planar graphs of the $\phi^3$ type: each propagator has one of 3
colors and all three propagators meeting at any vertex have different
colors. This 3-coloring problem can be formulated as the following
matrix model of three hermitean matrices $B$(lue), $W$(hite) and
$R$(ed)
 with the
action:
\eqn\trecol{S=N\Tr(g BWR + g BRW + B^2+W^2+R^2)}
To see that it coincides with the original Hoppe's model it is enough
to change the sign of one of 2 cubic terms (which is equivalent to the
change of anticommutator of $W,R$ to the commutator) and integrate
over one of the matrices, say, $B$. 
The critical behaviour (thermodynamical
limit) is due to the dominance of graphs of infinite size, which
renders the $g$-expansion of the partition function divergent and is
therefore determined by the singularity in $g^2$ closest to the
origin.  The latter is a solution of the equation
$g^{\prime}(m)=0$. It should appear for negative $g^2$ and corresponds
to the situation where all three cuts are located on the real axis,
symmetrically with respect to the origin. When $g$ increases, the
end-points of the cuts get closer and a singularity occurs when they
touch one another.  One can show, using the symmetry \prop\ of $W(z)$,
that when $g^2$ is real and negative, the saddle-point equation \spee\
becomes identical to eq. (3.3) of ref.  \kristjansen .  It is known
\gaudin\ that the critical behaviour of the dense $O(1)$ model is in
the universality class of the pure $2d$ quantum gravity.  For example,
the one-point function behaves as
$$\nu \sim (g_c-g)^{3\over 2}.  $$ 
Here is the
explicit formula for $(g^2)^{\prime}$:
\eqn\drvt{{d g^2 \over dm}  = - 9 \pi^4  {\bf K}^{4}
{{{\vartheta} ( {\vartheta} - 1) ( \vartheta - 1 + m)}\over{m ( 1-
m)}} .} 

\item{3.}The last assertion can be partially confirmed by
the study
of the specific heat at the
fixed chemical potential. From the eq. \sltn\ we get the
critical point
$x_c = - \left( {1\over \pi^2} + {1\over 12} \right)$,
which can be
inserted into eqs. \ledlog--\logu\
to yield $g_c^2 = - {1\over{3 \pi^4}}$,  which is negative
indeed. Note, however, that  the corresponding
value of $\ve = 2.16$ is way larger than the leading log
approximation allows us
to see.

\newsec{On the correlation functions in
the $d=6,10$
cases and directions for the  future}

This section is devoted to a work in progress.   We sketch the
possible similar saddle-point approach to the $d=6$
integral. We also attempt a fermionic representation for
the $d=10$ integral.

\subsec{Saddle-point approach to $d=6$ integral }

We keep
the same notations
for the resolvent $F$ and density $\rho$. We set $\e_1 +
\e_2 = 1$,
$\e_1 = \b$, $\e_2 = \g$.
Equation \spee\ is replaced by:
\eqn\spesix{\eqalign{ {{x}\over{g^{2}}} = &\ \  {\hat F}(x) -
{\half}
\left( F(x+i\b) + F(x-i\b) \right) \cr
& \ \ - {\half} \left( F(x+i\g) + F(x-i\g) \right) + {\half}
\left(
F(x+i) + F(x-i) \right).\cr}}
Let us introduce the derived functions:
\eqn\drvd{\eqalign{f(x) & =  {{i x^2}\over{2 g^2 \g}} + F
( x + {i\b
\over 2}) - F ( x - {i\b \over 2})\cr
g (x) & = {1\over 2} \left( f ( x + {i\g \over 2} ) - f (
x - {i\g
\over 2})\right).\cr}}
The saddle-point equation \spesix\ is equivalent to
\eqn\speesix{g \left( x + {i \over 2}\right)  + g \left( x - { i  \over
2}\right) = 0.}
The function $g(x)$ has four cuts: at $x \in  {\pm}
{i\over 2}, {\pm
}{i\over 2} ( \b - \g) + (-a, + a)$,
it is real: $g(\zb) = \overline{g(z)}$ and it is purely
imaginary for
$z \in i \IR$. It would be
nice to guess the correct function from the stated
properties. We
plan to return to this problem in
the future.

\subsec{Fermionic representation for the $d=10$ integral}

We now proceed with the  fermionic
representation of
the $d=10$ integral \harsh. Unfortunately we were not
able to
find such a
representation for all values of $\e_1, \e_2, \e_3$. However,
let us consider the limit $\e_3 \to - \e_1, \e_4 \to - \e_2$.
At the same time we keep
$$
e^{\mu} ( \e_1 + \e_3) = e^{\bar\mu}$$
finite.
We claim that, in this limit,
\eqn\frmn{\CZ_{V, \mu, \e}^{d=10}  = \langle 0 \vert
e^{H[t]} e^{\bar
\Omega_{\mu}} \vert 0 \rangle, }
where
\eqn\ffrm{\bar\Omega_{\mu} = e^{\mu} \int_{-\infty}^{+\infty}
{{dz}\over{2\pi i}} : \psi ( z - a) \psi ( z+ a)
\psi^{*}( z - b)
\psi^{*}
(z + b) : }
with $a = \half ( \e_1 + \e_2) , b = \half ( \e_1 - \e_2 )$.
Now the relation between $V$ and $U$ is modified into:
\eqn\ptnls{V(z) = U(z + a) + U(z-a) - U(z + b) - U( z-b).}

As in the index computation it is possible that,
by making the appropriate mass perturbation, we reduce
the $d=10$ integral for the gauge group $U(N)$ to the
products of $d=4$ integrals for gauge groups
$U(n_{1}) \times \ldots \times U(n_{k})$ with
$N = \sum_{l=1}^{\infty} l n_{l} $.
Under this assumption:
\eqn\sixdix{
\CZ^{d=10}(\mu, \e, V) =
\prod_{l=1}^{\infty} \CZ^{d=4}(l \mu, \epsilon, V) =
\prod_{l=1}^{\infty} {\Det} ( I + e^{l \mu} K) = \Theta
(\mu, K).
}
The latter expression is very interesting, since it
possesses certain modular properties and allows one to
deduce the
large-$N$ asymptotics using very little information about the
operator $K$ itself:
\eqn\asmtth{
\Theta (\mu, K)
\sim \exp 2 \sqrt{N \vert {\rm Li}_{2} (-K)\vert} , }
where
\eqn\dlog{
{\rm Li}_{2}(-K) = \sum_{l=1}^{\infty} {1\over{l^{2}}} {\Tr}
(-K)^{l}.}

\newsec{Conclusions}

Here we summarize  the results of our computations.
The integral \prtn,  which is a cousin of \mint, 
is studied in two regimes: at fixed $N$ and at fixed
$\mu$. In the
first case we got the
large-$N$ asymptotics in the 't Hooft limit, see
\yofm\ and \onpnt. In the
second case we showed that
the grand partition function is a particular tau-function
of the KP
hierarchy. In particular,
we obtained  eq. \hpf\ for the specific heat $u =
-2\p^2_{\mu} \CF$
in the case of quadratic potential $ V \sim \l z^2$:
$$2 u_{\l} + {1\over \l} u +  u u_{\m} + {1\over 6} (u -
u_{\m\m})_{\m} = 0.$$ 
In the large-$\mu,
\lambda$ limit,
we obtained the
simple explicit formula for the specific heat as a  
function of $x
= \mu /\l$:
\eqn\finale{-{{\pi^2}\over{4}} u^2 + u + {1\over{12}} = x.}
We observed various similarities  to the properties of
supersymmetric
gauge theories
in four dimensions and we hope that our results will find
their place
in the study
of dynamics of $D$-particles in various dimensions.

\newsec{Acknowledgements}

The research of N.~N. was supported by the Harvard Society of Fellows,
partially by NSF under grant PHY-9802709, partially by RFFI under
grant 96-02-18046 and partially by grant 96-15-96455 for scientific
schools.  N.~N. is also grateful to LPTHE at the University of Paris
VI-VII, to the Aspen Center for Physics, and to N.~ and
M.~Chechelnitsky for hospitality at various stages of this work.
V.~K.~ is grateful to the Physics Department of the Oxford University,
where part of this work was done, for the hospitality.  V.K. and
I.K. thank M. Fukuma for stimulating discussions and for bringing in
our attention ref. \hoppe .

\appendix{A}{Determination of $\a_0$}
We now present a trick that allows us to find the value of the
unknown
coefficient $\alpha_0$ in \sltn. If $\alpha_0 \neq 0$ then for
large $x$ one has $u \sim \pm \sqrt{2x \over{\alpha_0}}$
and
\eqn\lrgx{\p_{\mu} \CF \sim \pm {{\sqrt{2}\mu^{3 \over
2}}\over{3\xi
\sqrt{\alpha_0}}}.}
Below, we  rescale $\l \to \l / 2$ to be in agreement with the
notations of
\oprii.

Let  $e^{-E_k}$ be  the eigenvalues of the integral  
operator $\CK$.
{}From the determinant representation \grptfn\  
of the
 partition function, it follows   that
$e^{\CF(\mu, \xi)} = \prod_{k} \left( 1 + e^{\mu - E_k}  
\right).$
Correspondingly, the mean value of the number of particles is
given by
\eqn\mnvl{\langle N \rangle \equiv  \p_{\mu} \CF = \sum_{k}
{1\over{1 +
e^{E_{k} - \mu}}}.}
We are interested in the limit where both $\mu$ and $\xi$ (and
therefore
$E_{k}$, see below) are very large, i.e. a kind of low-temperature
limit for a Fermi-gas with energy levels given by the
spectrum of the operator $\log \CK$. In the low-temperature
limit we simply need to count the number of energy levels
below the Fermi level $\mu$.

The eigenvalue problem for the operator $\CK$ is similar
(although
far from being equivalent in general) to the eigenvalue
problem of the particle of unit mass, which is confined to move
at the positive semi-axis $y > 0$ and subject to the
spike-like
potential
$$
U(z, t) = {{\xi}\over{2}} y\sum_{n \in \IZ} \delta
( t -
n), \quad
\xi > 0.
$$
The operator $\CK$ is to be compared with the operator
$\CU_{1}$
of the evolution during the unit imaginary time.
The latter can be easily diagonalized:
\eqn\spctr{\CU_{1} f_{E} = e^{-E} f_{E},}
with $f_{E}(y) = A (y - {{2\tilde
E}\over{\xi}})$ ,
$\tilde E = E + {\log} \sqrt{2\pi} + {{\xi^{2}}\over{48}}$,
and $A(y)$ is the modified Airy function
\eqn\mdfa{A(y) = \int_{\gamma}
{{dp}\over{2\pi}} e^{ip y + {{p^{2}}\over{4}} + i
{{p^{3}}\over{3\xi}}}}
where the contour $\gamma$ is such that ${\Im}p^{3} > 0$ as
$p \to \infty$ along $\gamma$.
The spectrum  is determined from the condition that
$f_{E}(0) = 0$,
i.e.
$A(- {{2\tilde E}\over{\xi}}) = 0$.
For large values of $E$ this equation can be solved
using quasi-classics. It gives
\eqn\asmpt{E_{k} \sim \half \left(
{{3\pi \xi k  }\over{2}} \right)^{2 \over 3}, \qquad
k \to \infty. }
Assuming that for our problem we may use the same
asymptotics we
conclude that
\eqn\mnvln{\langle N \rangle \sim {{2^{5 \over
2}}\over{3\pi\xi}}
{{\mu^{3 \over 2}}},}
which means that
\eqn\fnl{\alpha_0 = {{\pi^{2}}\over{2}}.}

{\it Remark.} It is interesting to note that a similar
Schr\"odinger
problem arises in the Born-Oppenheimer
approximation to the quantum mechanics of a particle in two
dimensions
confined by the potential $x^2 y^2$, which is a good
model for the
matrix potential ${\Tr} [ X, Y]^2$, see \hoppe.

\appendix{B}{Solution of the one-matrix model from
the KP equation and double scaling limit.}

As an example illustrating  the application of the KP
hierarchy to matrix integrals, we will derive the critical
singularity and the double scaling limit of the
one-matrix integral:
\eqn\ommo{
Z_{N} ( t) = \int {\CD} M \exp N {\rm Tr} \sum_{q=1}^{3}
t_{q} M^{q}}
in the case of a cubic potential.
The fact that the  matrix integral \ommo\ is a tau-function of the KP 
hierarchy has been established in \morozov . A direct derivation of the 
Hirota equations
\Hiilr\ from the matrix integral can be found 
in \IvanRazl . 
The free energy
\eqn\fre{
F_{N} = {1\over{N^{2}}} {\rm log} Z_{N}}
and  the specific heat
\eqn\sphe{u ( t) = 2 \partial_{t_{1}}^{2} F_{N}}
can be obtained from the  KP equation \KPe,   
if we take it into account that  the specific heat 
is actually a function of a single parameter:
\eqn\scl{u = t^{-{2\over 3}}_{3}
\psi \left({{t_{1}}\over{t_{3}^{1\over
3} }} - {{t_{2}^{2}}\over{3t_{3}^{4\over 3}}} \right)}
and is given in the Gaussian limit by
\eqn\gsli{
t_{3} = 0 : \quad u = - {1\over{t_{2}}}.}
By inserting  \scl\ into \KPe\  we derive that
the function $f$ obeys the following ordinary
differential equation
\eqn\ode{
\psi + 2x \psi^{\prime} + 9 \psi\psi^{\prime} +  
{3\over{2N^{2}}}
\psi^{\prime\prime\prime} = \psi_{0}}
where $\psi_{0}$ is a constant,
\eqn\xf{
x = {{t_{1}}\over{t_{3}^{1\over 3} }} -
{{t_{2}^{2}}\over{3t_{3}^{4\over 3}}}, \quad
\psi = t_{3}^{2\over 3} u.}
In the large-$N$ limit, we have
an algebraic  equation
for $x(\psi)$:
\eqn\lrnl{
(\psi- \psi_{0}) {{dx}\over{d\psi}} + 2x + 9\psi = 0,}
whose solution depends on two constants $(\psi_{0}, \b)$:
\eqn\gnsl{
x = {{\b}\over{(\psi- \psi_{0})^{2}}} - 3\left(\psi -  
{1\over{2}}
\psi_{0}\right). }
By comparing this with  the Gaussian limit, we get: $\psi_{0}=0,
\b=-{1\over 3}$. Hence the large-$N$ result is:
\eqn\thlrg{
t_{1} t_{3} - {1\over{3}} t_{2}^{2} = - 3 t_{3}^{2} u -
{1\over{3u^{2}}}.}
The critical point is
\eqn\crtp{
x_{c} = - \left({9\over{2}} \right)^{2\over 3} , \quad  
\psi_{c} =
\left({2\over{9}}\right)^{1\over 3}.}

To compare with the results of Brezin et al.  \BIPZ\ we set:
$$
t_{1} = 0, \quad t_{2} = {1\over{2N}}, \quad t_{3} =
{{g}\over{N^{3\over 2}}}
$$
and get $x = -{1\over{12 g^{4\over 3}}}$. Hence
$$
g_{c}^{2} = {1\over{108\sqrt{3}},}
$$
which is exactly the result of the tedious computations \BIPZ\
using the distribution of the eigenvalues in the large-$N$ limit.
Equation  \KPe\  also allows a trivial derivation of the
double scaling
limit
for the pure $2d$ quantum gravity \dsl.
This limit consists in sending
$x-x_c$ to zero and $N$ to infinity in such a way that
the double
scaling variable $z=N^b (x-x_c)$ remains finite. We try the
 ansatz:
$\psi = \psi_c+N^a \up (z).$
Since  $\psi \simeq \psi_c + {\rm const} \cdot   
\sqrt{x-x_c}$ we
have:  $b= - 2a$.
We then plug this ansatz into \KPe\ and obtain
\eqn\dbl{\up^{\prime\prime\prime} +6N^{2+5a}\up
\up^{\prime}+{2\over 3} \psi_c N^{2+5a}
+\left({4\over 3} z \up^{\prime} +{2\over 3} \up\right) N^{2 + 6a} =0.}
To keep here the non-linear term (the source of all higher-genus
corrections  $\sim 1/N^{2g}$) we  impose the condition:
$2+ 5a=0$,
which gives
$$a=-{2\over 5}, \, \, \, b={4\over 5}.$$
 One immediately sees that
the last term in \dbl\ vanishes in the double scaling
limit ($N^{2+  6a} = N^{-2\over 5} \to 0 $). Integrating 
\dbl\ once
with
respect to $z$,
we finally obtain the  Painlev\'e $\II$ equation:
\eqn\pnlv{\up^{\prime\prime}+3\up^2= c z,}
where $c=-{2\over 3} \psi_c = - {2^{4\over 3} \over  
3^{5\over 3}}$.

\noindent{\it Remark.} We obtained this equation without
making  use at all of
the method of orthogonal polynomials --  the only 
technique known
until now for these purposes. This  raises the  hope that the
method of
Hirota
equations proposed here will allow us to obtain the
non-perturbative
description
(beyond the loop expansion) of
some interesting non-critical string theories, such as           
$O(2)$ model   \kostovOn\ whose grand partition
is known to satisfy the KdV hierarchy \kostovOnKdV .

\listrefs
\bye